\documentclass[11pt]{article}
\pdfoutput=1
\usepackage{jcappub}
\usepackage{graphicx,amssymb,amsmath,xcolor}
\graphicspath{{figs/}}
\newcommand{\figh}[2]{\includegraphics[height=#1\textheight]{#2}}
\newcommand{\figw}[2]{\includegraphics[width=#1\textwidth]{#2}}

\newcommand{\Cl}{\ensuremath{C_\ell}}
\newcommand{\lmax}{\ensuremath{\ell_\mathrm{max}}}
\newcommand{\Neecr}{\ensuremath{N_\mathrm{UHECR}}}
\newcommand{\Diso}{\ensuremath{D(\mathrm{iso})}}
\newcommand{\Dmix}{\ensuremath{D(\mathrm{mix})}}
\newcommand{\keuso}{\mbox{K-EUSO}}

\begin{document}

\title{%
	 Prospects of detecting a large-scale anisotropy of ultra-high-energy
	 cosmic rays from a nearby source with the K-EUSO orbital telescope}

\author[a]{Oleg Kalashev,}
\author[a,b,c]{Maxim~Pshirkov,}
\author[d]{Mikhail~Zotov}

\affiliation[a]{Institute for Nuclear Research of the Russian Academy of
	Sciences, Moscow, 117312, Russia}
\affiliation[b]{Sternberg Astronomical Institute, Lomonosov Moscow State
	University, Moscow, 119992, Russia}
\affiliation[c]{Lebedev Physical Institute, Pushchino Radio
	Astronomy Observatory, 142290, Russia}
\affiliation[d]{Skobeltsyn Institute of Nuclear Physics,
	Lomonosov Moscow State University, Moscow, 119991, Russia}

\abstract{
	KLYPVE-EUSO (K-EUSO) is a planned orbital detector of ultra-high-energy
	cosmic rays (UHECRs), which is to be deployed on board the
	International Space Station.  K-EUSO is expected to have a
	uniform exposure over the celestial sphere and register from 120 to
	500 UHECRs at energies above 57~EeV in a 2-year mission.  We employed
	the TransportCR and \mbox{CRPropa}~3 packages to estimate prospects of
	detecting a large-scale anisotropy of ultra-high-energy
	cosmic rays from a nearby source with K-EUSO.
	Nearby active galactic nuclei Centaurus~A,
	M82, NGC~253, M87 and Fornax~A were
	considered as possible sources of UHECRs.
	A minimal model for extragalactic cosmic rays and neutrinos
	by Kachelrie\ss, Kalashev, Ostapchenko and
	Semikoz (2017) was chosen for definiteness.
	We demonstrate that an observation of $\gtrsim300$ events will allow
	detecting a large-scale anisotropy with a high confidence level
	providing the fraction of from-source events is $\simeq$10--15\%,
	depending on a particular source.
	The threshold fraction decreases with an increasing sample size.
	We also discuss if an overdensity originating from a nearby source
	can be observed at around the ankle in case a similar anisotropy
	is found beyond 57~EeV.
	The results are generic and hold for other future experiments
	with a uniform exposure of the celestial sphere.

}
\emailAdd{kalashev@inr.ac.ru}
\emailAdd{pshirkov@sai.msu.ru}
\emailAdd{zotov@eas.sinp.msu.ru}

\arxivnumber{1810.02284}
\maketitle

\section*{Introduction}

Ultra-high-energy cosmic rays (UHECRs) with energies above $\sim50$~EeV,
sometimes called extreme-energy cosmic rays, were first registered
almost 60 years ago~\cite{1961PhRvL...6..485L} but their nature and
sources still remain an open problem of astrophysics and cosmic ray
physics.  UHECRs are supposed to be produced in extragalactic sources
and this has been recently corroborated by an observational finding of
the dipole anisotropy in the arrival directions of UHECRs with energies
above 8~EeV~\cite{Auger-dipole-2017}.  Different classes of
astrophysical objects  are considered as possible sources of UHECRs,
among them gamma-ray
bursts~\cite{PhysRevLett.75.386,1995ApJ...449L..37M,1995ApJ...453..883V},
young millisecond pulsars and
magnetars~\cite{2000ApJ...533L.123B,2003ApJ...589..871A,2012ApJ...750..118F},
tidal disruption
events~\cite{2009ApJ...693..329F,2012EPJWC..3907005F,2018A&A...616A.179G},
active galactic nuclei (AGN) of different
types~\cite{2009PhRvD..80l3018P} and mechanisms of acceleration,
including blazars~\cite{2012ApJ...749...63M,2018arXiv180401064N}, black hole
jets~\cite{2009NJPh...11f5016D,2018NatPh..14..396F,2018PhRvD..97b3026K}
and several other, see, e.g.,~\cite{2011ARA&A..49..119K} for a more
in-depth discussion. 

In this work, we focus on possible large-scale anisotropy signatures of
a model by Kachelrie\ss, Kalashev, Ostapchenko and
Semikoz~\cite{Kachelriess:2017tvs} (KKOS in what follows).
We use the model as a benchmark scenario which
represents a much broader class of models with strong nearby UHECR
sources and an intermediate composition, rather than something unique.
Although some particular features of other models can vary, we expect
our estimations might be true for them as well.

The KKOS model can explain the observed energy spectrum and mass composition of
cosmic rays (CR) with energies above $\sim10^{17}$~eV, and
matches the high-energy neutrino flux measured by IceCube.  The scenario
assumes that UHECRs are produced by (possibly a subclass of) AGN. The
model does not rely on any particular acceleration mechanism, although
it was shown that UHECR production could proceed either via shock
acceleration in accretion shocks~\cite{1986ApJ...304..178K} or via
acceleration in regular
fields~\cite{Blandford76,Lovelace1976,Blandford:1977ds,MacDonald:1982zz,Neronov:2007mh}
close to a supermassive black hole (SMBH). Alternatively, UHECRs could
be produced in large-scale radio jets~\cite{Rachen:1992pg} or via a
two-step acceleration process in the jet and a radio-lobe
\cite{Hardcastle:2008jw}.  The model neglects the acceleration process
details and just relies on the following basic assumptions: (i)~the
energy spectra of nuclei after the acceleration phase follow a power-law
with a rigidity dependent cutoff
\[
	j_\mathrm{inj}(E) \propto E^{-\alpha} \exp[-E/(ZE_{\max})]\,; 
\]
(ii) the CR nuclei diffuse first through a zone dominated by
photo-hadronic interactions, and then they escape into a second zone
dominated by hadronic interactions with gas.  It was shown that a good
fit to the CR energy spectrum can be obtained assuming only hadronic
interactions of UHECRs with gas around their sources, but it was
difficult to reproduce the observed distribution of~$X_{\max}$ of
CR-induced air showers.  Only after adding photo-nuclear interactions
with a relatively large interaction depth, which suggests that UHECRs
are accelerated close to SMBHs, it was possible to reduce significantly
the fraction of heavy nuclei in the primary fluxes and therefore fit
satisfactorily both the spectrum and composition data on UHECRs.
Moreover, the secondary high-energy neutrino flux obtained in the
scenario matches the IceCube measurements~\cite{Aartsen:2016xlq}, while
the contribution of unresolved UHECR sources to the extragalactic
$\gamma$-ray background~\cite{Ackermann:2014usa} is of the order
of~30\%.

\begin{figure}[!ht]
	\centering
	\figw{.49}{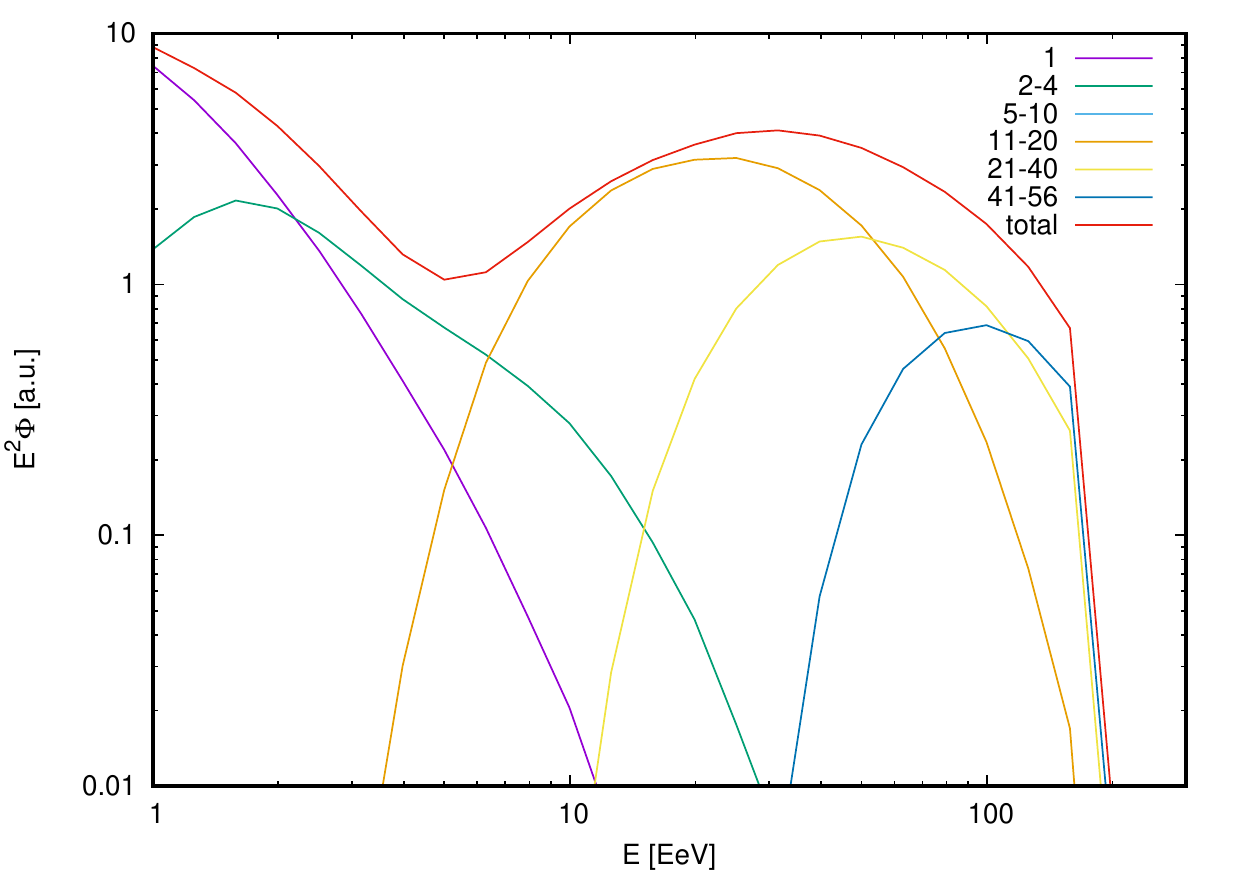}
	\figw{.49}{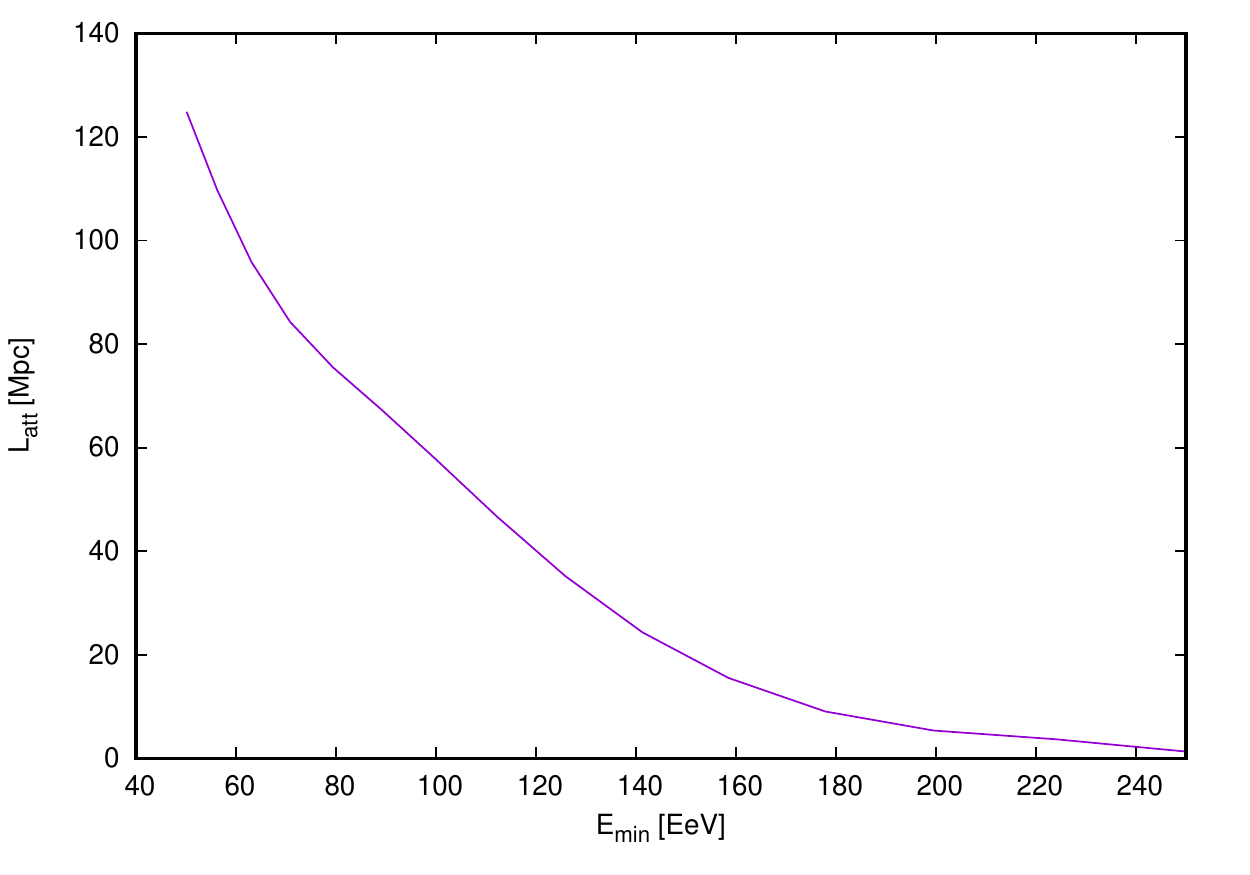}
	\caption{The effective CR source energy spectrum for different mass
	components (left) and integral flux above $E_\text{min}$ suppression
	length (right) in the KKOS model.}
	\label{fig:spectrum}
\end{figure}

The spectrum of CR leaving the source environment (i.e., after passing
the second zone) in the best fit obtained with the KKOS model for $\alpha=1.5$ is shown
in the left panel of Fig.~\ref{fig:spectrum} for several mass ranges. In the
right panel, we illustrate the attenuation effect on the integral flux by
plotting the distance at which the total flux above the given energy drops
by factor of $1/e$. Note that the integral flux attenuation length
depends both on the initial source spectrum and composition.%
\footnote{For the photodisintegration process in general, the energy
loss length of the leading nucleus with atomic number~$A$ and
Lorentz-factor $\gamma$ can be roughly related to its interaction length
$L_\mathrm{int}(A,\gamma)$ as $L_\mathrm{loss}\simeq
L_\mathrm{int}\times \delta A/A \simeq  L_\mathrm{int}/A$, where $\delta
A\simeq1$ is the average number of nucleons lost by the nucleus in a
single interaction. Since $L_\mathrm{int}$ moderately depends on~$A$
while $\gamma$ remains approximately constant in the chain of
interactions, $L_\mathrm{loss}$ does not change dramatically along
the particle trajectory until $A=1$. This is not the case for protons:
$L_{\mathrm{loss},p}$ drops as soon as the GZK cut-off energy is
reached.}

Due to strong propagation effects, the observable cosmic ray mass
composition and energy spectrum can vary considerably for each
individual source even though the injection spectrum is precisely the
same for all of them. Moreover, at energies $E \simeq 150$~EeV, where
the attenuation length for the integral flux drops to tens of Mpc, the
CR flux may be dominated by the contribution of a nearby source located
within 20~Mpc from the Milky Way.  This, in turn, can lead to a
substantial large-scale anisotropy in the UHECR flux. Orbital detectors
with a sufficiently large exposure will provide good opportunities for
studying this effect due to their possibly almost uniform exposure of
the whole celestial
sphere~\cite{2014A&A...567A..81R,2015JCAP...01..030O,2015JHEAp...8....1D}.

One of the future orbital experiments that are being actively developed
today is the KLYPVE-EUSO (\keuso) telescope, which is aimed to be
installed on board the International Space Station (ISS) in 2022 for a
2-year
mission~\cite{klypve-2015bullras,k-euso-ICRC2015,k-euso-ICRC2017,k-euso-ICRC2017-science}.
\keuso{} is a further development of a technique of registering UHECRs
via ultra-violet radiation emitted by extensive air showers in the
atmosphere of Earth from a low-orbit satellite, implemented for the
first time in the TUS detector~\cite{tus-ssr-2017,tus-jcap-2017}.  The
telescope is expected to have a Schmidt-type optical system with the
main mirror-reflector of a 4~m diameter, an entrance pupil of a 2.5~m
diameter and a 1.7~m focal length. A round-shaped field of view of
$40^\circ$ will provide an instantaneous geometrical area of nearly
$6.7\times10^4$~km$^2$ at sea level for the ISS altitude around 400~km.
It is expected that \keuso{} will register from 120 up to almost 500
UHECRs with energies above 57~EeV in two years.  The difference between
the lower and the upper boundaries of the estimate arises from the
difference in the energy spectra of the Pierre Auger Observatory and the
Telescope Array~\cite{k-euso-ICRC2017-science}.
Capabilities of the previous version of \keuso\ (KLYPVE) to detect
the Telescope Array hotspot were studied earlier~\cite{STZ-2016}.

In what follows, we study if \keuso{} will be able to detect a
large-scale anisotropy of arrival directions of UHECRs with energies
above 57~EeV originating in a nearby AGN in the KKOS model.  However,
the presented results are valid for any other future experiment with a
uniform exposure of the celestial sphere, including the POEMMA
mission~\cite{POEMMA}.\footnote{%
	Possible signatures of a large-scale anisotropy of 
	UHECRs arriving from a single nearby source has already
	been studied by different authors, see, e.g.,~\cite{2016PhRvD..93f3002H,2019JCAP...01..018D}.
	This work employs a different model of CR sources and another
	technique of identifying an anisotropy.}

\section{Method of the analysis}

In what follows, we consider five possible sources of UHECRs that are
often discussed in literature and satisfy the KKOS model, namely,
NGC~253, Centaurus~A, M82, M87 and Fornax~A.  These are radio-loud
galaxies located at distances~$d\approx3.5\dots20$~Mpc from the Milky
Way.  For each source located at a given distance~$d$, we calculated
the energy spectrum and mass composition of the CR flux crossing the
Milky Way boundary using a public numerical code
TransportCR~\cite{transportcr}, which was also used in the original
work~\cite{Kachelriess:2017tvs}.  A contribution of other sources was
approximated by an isotropic component.  The spectrum and mass
composition for the isotropic component was calculated by solving the
transport equation with a homogeneous source distribution for distances
$R>2^{1/3}d$, and zero density for smaller distances.

We assumed that deflections of CR nuclei in the inter-galactic space are
small, so that nuclei accelerated at a particular source arrive to the
Milky Way within $1^\circ$  from the actual direction to the source.
The \mbox{CRPropa}~3 package~\cite{CRPropa} (GitHub snapshot of 24th June,
2018) with the Jansson--Farrar model of the
Galactic magnetic field (GMF)~\cite{JF12}
was employed to calculate deflections of nuclei in the Galaxy on
their way to the Solar system.
All three components of the
GMF present in the model (the regular, striated and turbulent ones) were
utilised in simulations.  Backtracking was performed for all possible
pairs $(Z,E)$ in the spectra to obtain maps of apparent arrival
directions of different nuclei to Earth given original directions of
approaching the Milky Way. Calculations were made on the
HEALPix\footnote{\url{https://healpix.sourceforge.io}} grid with
$N_\mathrm{side}=512$ providing an angular resolution of the order of~$7'$.
This is far beyond the angular resolution of the \keuso\ experiment.
It was chosen to reliably recover an observed distribution of
arrival directions of nuclei after their propagation in the Galactic
magnetic field.

There are a number of mathematical tools traditionally used for studying
the large-scale anisotropy of cosmic rays. Historically the first one is
the harmonic analysis in right ascension, see~\cite{Auger-dipole-2017} for
a recent application. 
In practice, the effectiveness of the method is mostly confined
to the lowest multipoles due to the small statistics of UHECRs.
Another approach is based on calculating the angular power spectrum,
see, e.g., \cite{Sommers-2001,power_spectrum}.
We performed the respective analysis by preparing maps of the relative
intensity of the CR flux
\begin{equation}\label{eq:flux}
	\delta I_i = \frac{N_i - \langle N\rangle_i}{\langle N\rangle_i},
\end{equation}
where $N_i$ and $\langle N\rangle_i$ are the number of ``observed''
events and the number of ``reference'' events assuming the isotropic
flux in the $i$th pixel of the HEALPix map.
Coefficients of the angular power spectrum
\begin{equation}\label{eq:Cl} 
    \Cl = \frac1{2\ell+1}\sum_{m=-\ell}^{+\ell}|a_{\ell m}|^2
\end{equation}
were calculated using the \texttt{anafast} program of HEALPix.
Coefficients $a_{\ell m}$ in Eq.~(\ref{eq:Cl}) are the multipolar
moments of the spherical harmonics used to decompose
the relative intensity~$\delta I_i$, defined in Eq.~(\ref{eq:flux}).
Notice the coefficient $C_0=0$ in our case since~$\delta I_i$ does not
include the angular average.

The IceCube and Auger experiments suggested a simple approach
that allows estimating the total impact of different multipoles
in deviating from an isotropic distribution of arrival directions
and also penalises statistically the search over many angular
scales~\cite{IceCube-aniso-2007,Auger-aniso-2017}.
They suggested to calculate an estimator
\begin{equation}
	D^2(\text{sample}) = \frac{1}{\lmax}
			\sum_{\ell=1}^{\lmax}
			\left(\frac{C_{\ell,\mathrm{sample}}
				-\langle C_{\ell,\mathrm{iso}}\rangle}
				{\sigma_{\ell,\mathrm{iso}}}\right)^2,
	\label{eq:D2}
\end{equation}
where ``sample'' is either ``data'' when applied to experimental data
or ``iso'' when applied to estimate the deviation of one isotropic sample
from an averaged isotropic flux.
Variables
$C_{\ell,\mathrm{sample}}$, $\langle C_{\ell,\mathrm{iso}}\rangle$
and $\sigma_{\ell,\mathrm{iso}}$ are, respectively, the~\Cl{} observed
in the sample (either ``data'' or ``iso''),
and the average and the standard deviation of~\Cl{} for isotropic
expectations, all of them calculated at a given scale~$\ell$. In
practice, $\langle C_{\ell,\mathrm{iso}}\rangle$ and
$\sigma_{\ell,\mathrm{iso}}$ are evaluated using a simulated isotropic
flux with the same number of events and exposure as for the
data~\cite{Auger-aniso-2017}.  One can choose a certain confidence level
for defining a threshold to accept or reject the isotropy hypothesis and
then compare a value of~$D^2(\mathrm{data})$ calculated for the data to
the distribution of~$D^2(\mathrm{iso})$ obtained for the isotropic flux.

We tried this approach but found another function to be slightly
more sensitive to deviations from an isotropic distribution than
the one given by Eq.~(\ref{eq:D2}).
Namely, all results presented below are based on calculating an estimator
\begin{equation}
	D(\text{sample}) = \frac{1}{\lmax}
		\sum_{\ell=1}^{\lmax}
		\frac{C_{\ell,\mathrm{sample}}
			-\langle C_{\ell,\mathrm{iso}}\rangle}
			{\sigma_{\ell,\mathrm{iso}}},
	\label{eq:D}
\end{equation}
which is the same as the one used by the Pierre Auger Collaboration
but without the square of the summands. This allows taking into
account the fact that the expected deviations from the isotropic
case are one-sided (positive).
Since we are using simulated data instead of experimental results,
the $C_{\ell,\mathrm{data}}$ coefficients in Eq.~(\ref{eq:D2}) are
to be replaced with~$C_{\ell,\mathrm{mix}}$, which denote~\Cl{}
obtained for a simulated mixture of an isotropic flux and
cosmic rays arriving from a particular source.
Contrary to the case when one employs
experimental data, we needed to simulate many mixed samples to obtain
the distribution of~\Dmix{} for each source.

\section{Main results}

Let us consider the simplest case of a large-scale anisotropy arising
from an impact of a single source.  
Both~\Diso\ and~\Dmix\ are random variables in our case, thus one
needs to compare their distributions.
As the null hypothesis, we assume that arrival directions of
a mixed sample of UHECRs obey an isotropic distribution.
We adopted the value of the error of the second kind $\beta=0.05$
and searched for a fraction $F_1/F_\mathrm{tot}$
of from-source events in the total flux such that the
error of the first kind $\alpha\lesssim0.01$.\footnote{%
    Let us remind that an error of the first kind (a type~I error)
    stands for false positive errors, i.e., the rejection of a true
    null hypothesis,
    while an error of the second kind (a type~II error) is committed
    when a false null hypothesis is not rejected.}
We performed simulations for
$\Neecr=100, 200,\dots, 500$ to cover the whole possible range
of UHECRs to be detected by \keuso{} above 57~EeV.
The main results are presented in
Tables~\ref{table:percent} and~\ref{table:alpha}.

The numbers in Table~\ref{table:percent} give the percentage of events
arriving from a particular source in a sample of size~\Neecr, such that
the above condition is satisfied ($\beta=0.05$, $\alpha\lesssim0.01$).
Table~\ref{table:alpha} provides actual values of~$\alpha$ found in
each case.
For example, the error of the
first kind $\alpha\approx0.004$ as soon as the fraction of
events arriving from Cen~A in the otherwise isotropic sample
of the size $\Neecr=500$ is $\ge9\%$.
Thus, Table~\ref{table:percent} provides the percentage of from-source
events in the whole sample that will allow detecting a large-scale
anisotropy of UHECRs arriving from a particular source with a
sufficiently small error of the first kind.

\begin{table}[!ht]
	\caption{Percentage of UHECRs arriving from five candidate sources
	in samples of sizes $\Neecr=100,\dots,500$ such that the error of the first kind
	$\alpha\lesssim0.01$ for the null hypothesis of isotropy
	providing the second kind error $\beta=0.05$.
	The accuracy of the numbers is~$\pm1$.
	}
	\label{table:percent}
	\medskip
	\centering
	\begin{tabular}{|l|c|c|c|c|c|}
		\hline
		\Neecr    &   100 &   200 &   300 &   400 &   500\\
		\hline
		NGC 253   &    17 &    12 &    10 &     8 &     7\\
		Cen A     &    21 &    14 &    12 &    10 &     9\\
		M82       &    26 &    18 &    14 &    12 &    11\\
		M87       &    29 &    20 &    16 &    14 &    12\\
		Fornax A  &    19 &    13 &    11 &     9 &     8\\
		\hline
	\end{tabular}
\end{table}

\begin{table}[!ht]
	\caption{Probabilities of the first kind errors~$\alpha$ for
	the percentage of from-source events given in
	Table~\ref{table:percent}.
	}
	\label{table:alpha}
	\medskip
	\centering
	\begin{tabular}{|l|c|c|c|c|c|}
		\hline
		\Neecr    &   100 &   200 &   300 &   400 &   500\\
		\hline
		NGC 253   & 0.007 & 0.004 & 0.002 & 0.008 & 0.010\\
		Cen A     & 0.006 & 0.010 & 0.003 & 0.005 & 0.004\\
		M82       & 0.009 & 0.006 & 0.009 & 0.008 & 0.003\\
		M87       & 0.006 & 0.005 & 0.005 & 0.003 & 0.007\\
		Fornax A  & 0.007 & 0.009 & 0.003 & 0.009 & 0.009\\
		\hline
	\end{tabular}
\end{table}

\begin{figure}[!ht]
	\centering
	\figh{.26}{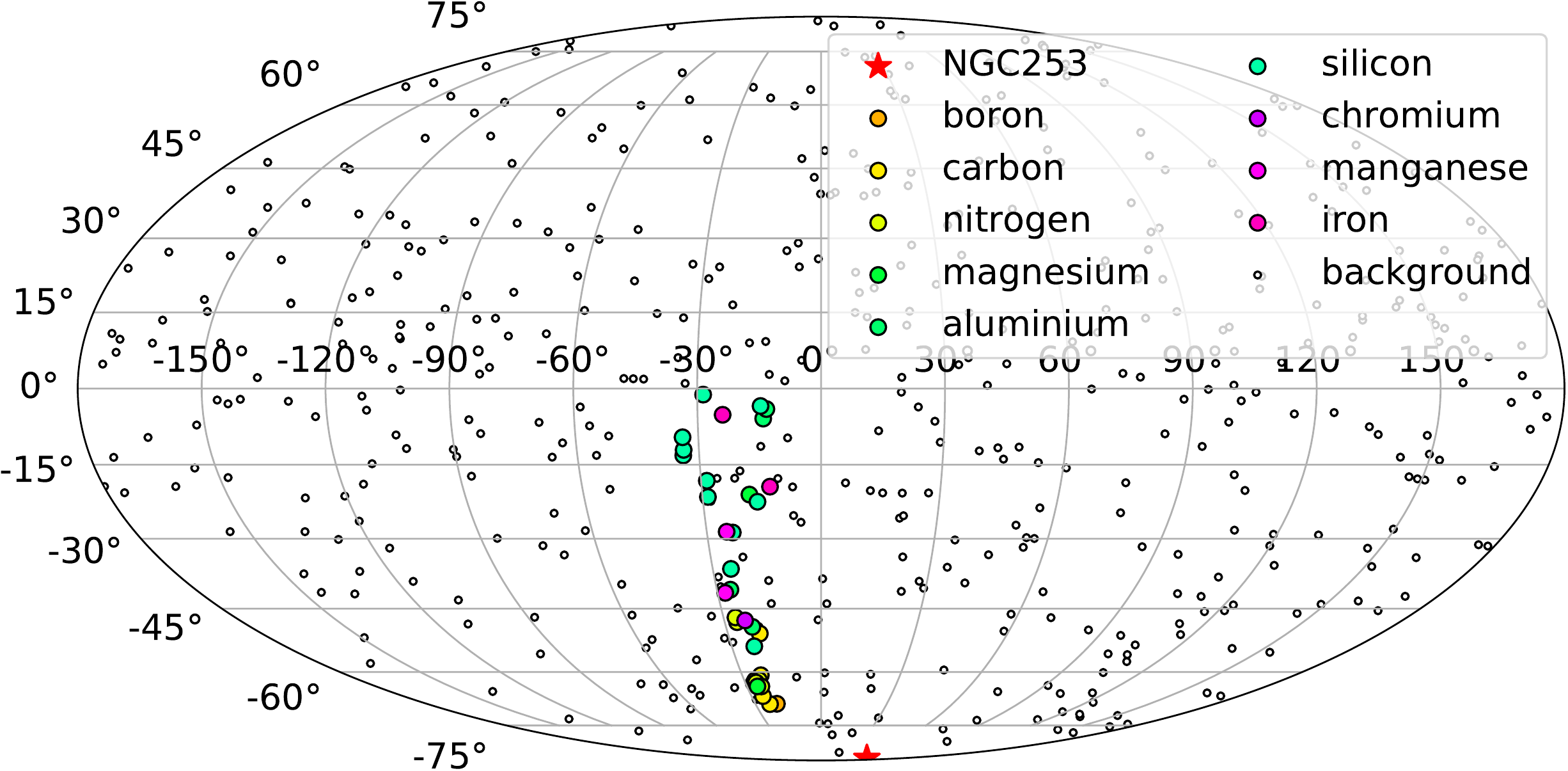}\\
	\figh{.24}{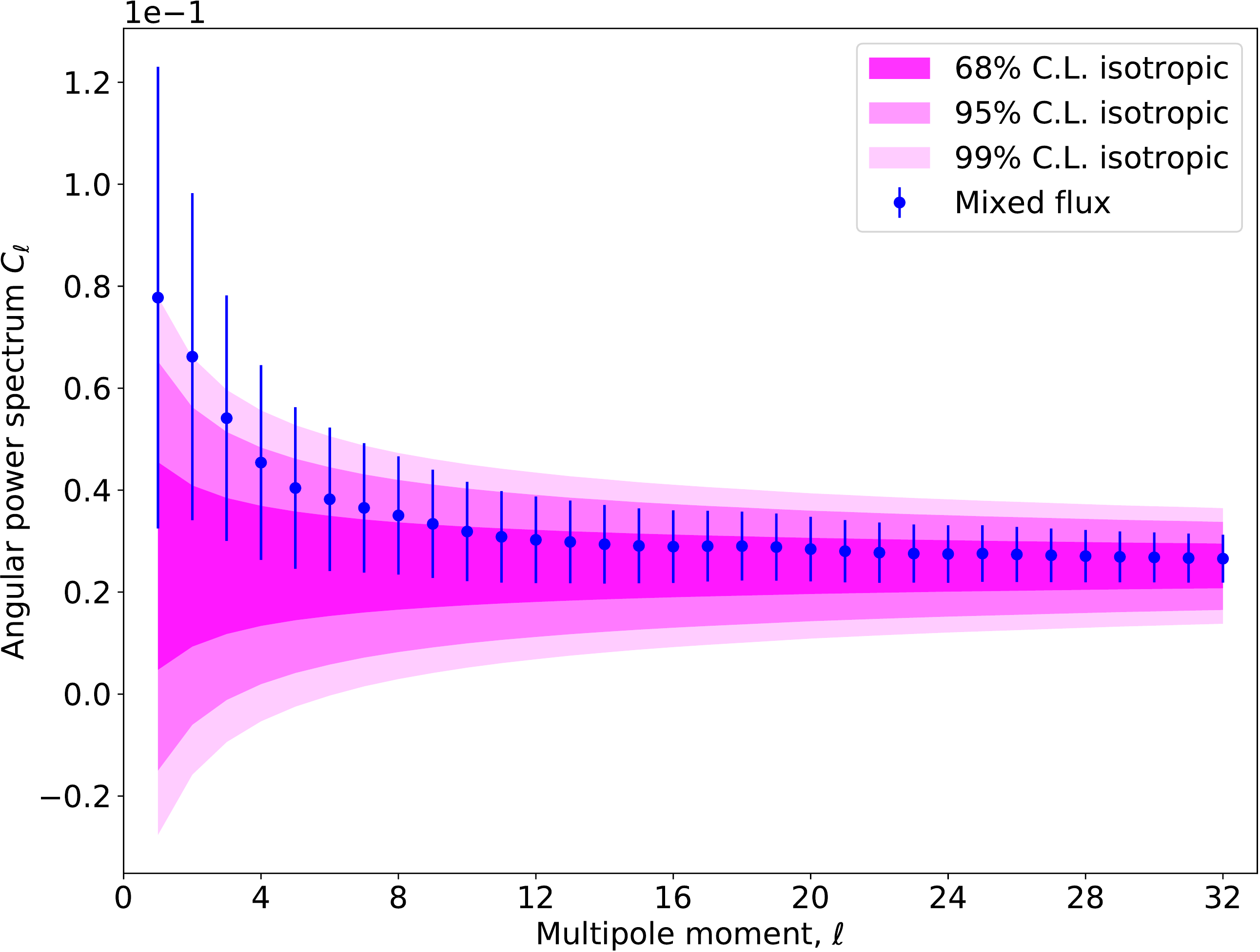}
	\figh{.232}{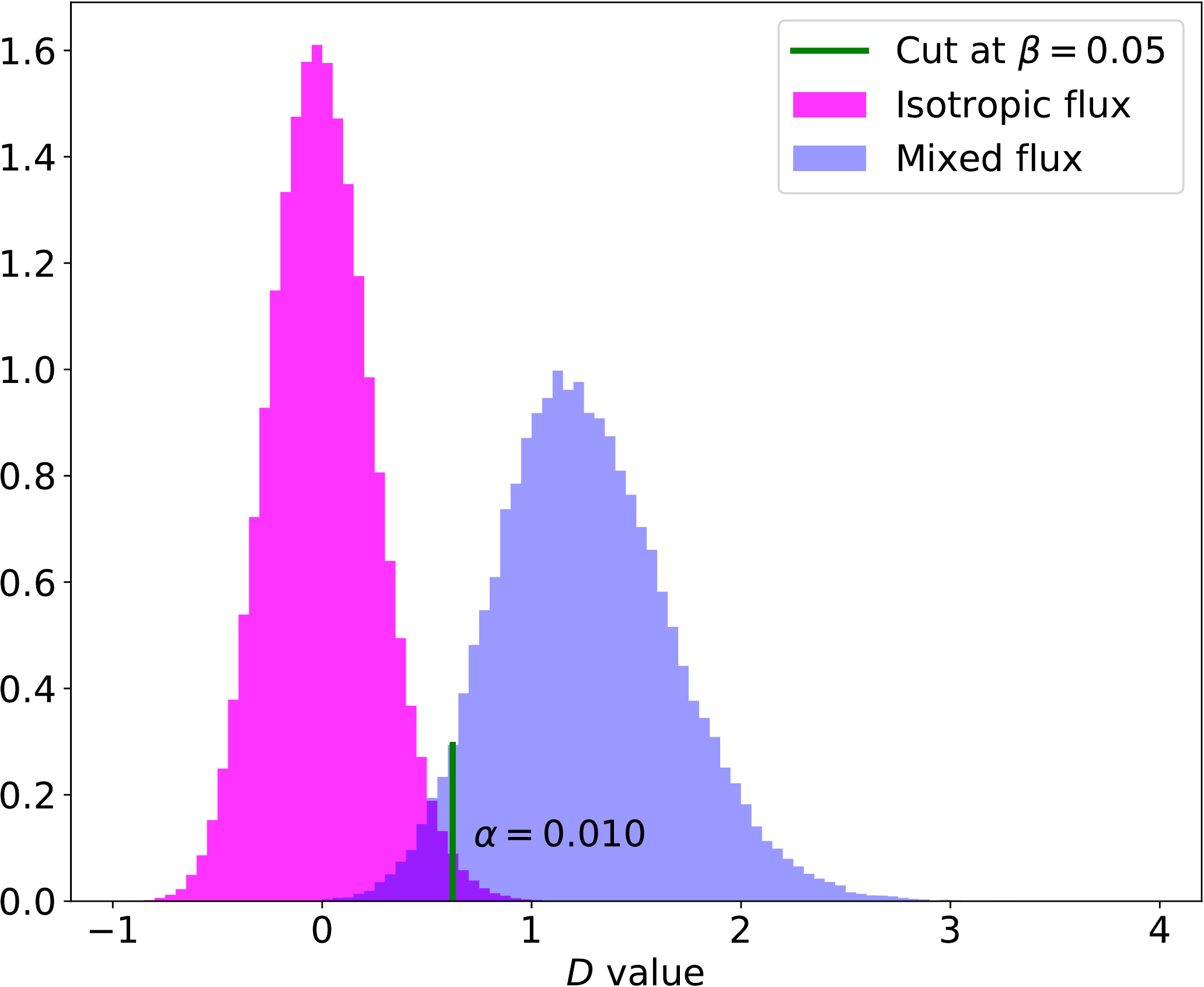}
	\caption{The case of $\Neecr=500$ and 7\% events coming from NGC~253.
	Top panel: an example of  a possible pattern of arrival directions.
	UHECRs that form an isotropic background are shown with small
	open circles.
	UHECRs arriving from the source are shown with coloured circles
	according to the type of the respective nucleus.
	The position of the source is indicated by the star.
	The map is shown in Galactic coordinates in the Mollweide
	projection.
	Left bottom panel: angular power spectrum~\Cl\
	for isotropic and mixed samples.
	Confidence intervals for 68\%, 95\% and 99\% levels are shown
	with different shades of magenta for the isotropic distribution.
	Blue dots with error bars indicate~\Cl\ for 50,000 mixed samples
	of size 500 with~7\% events in each sample coming from the source.
	Right bottom panel: the histograms show empirical probability
	distribution functions of~\Diso\ and~\Dmix\ calculated
	according to Eq.~(\ref{eq:D}).
	The vertical red line marks a value of $\Dmix$ such that only 5\% of
	all~\Dmix\ values are less than this particular value.
	This corresponds to the error of the second kind $\beta=0.05$.
	The error of the first kind~$\alpha$ is also indicated in the panel.
	}
	\label{fig:NGC253}
\end{figure}

\begin{figure}[!ht]
	\centering
	\figh{.21}{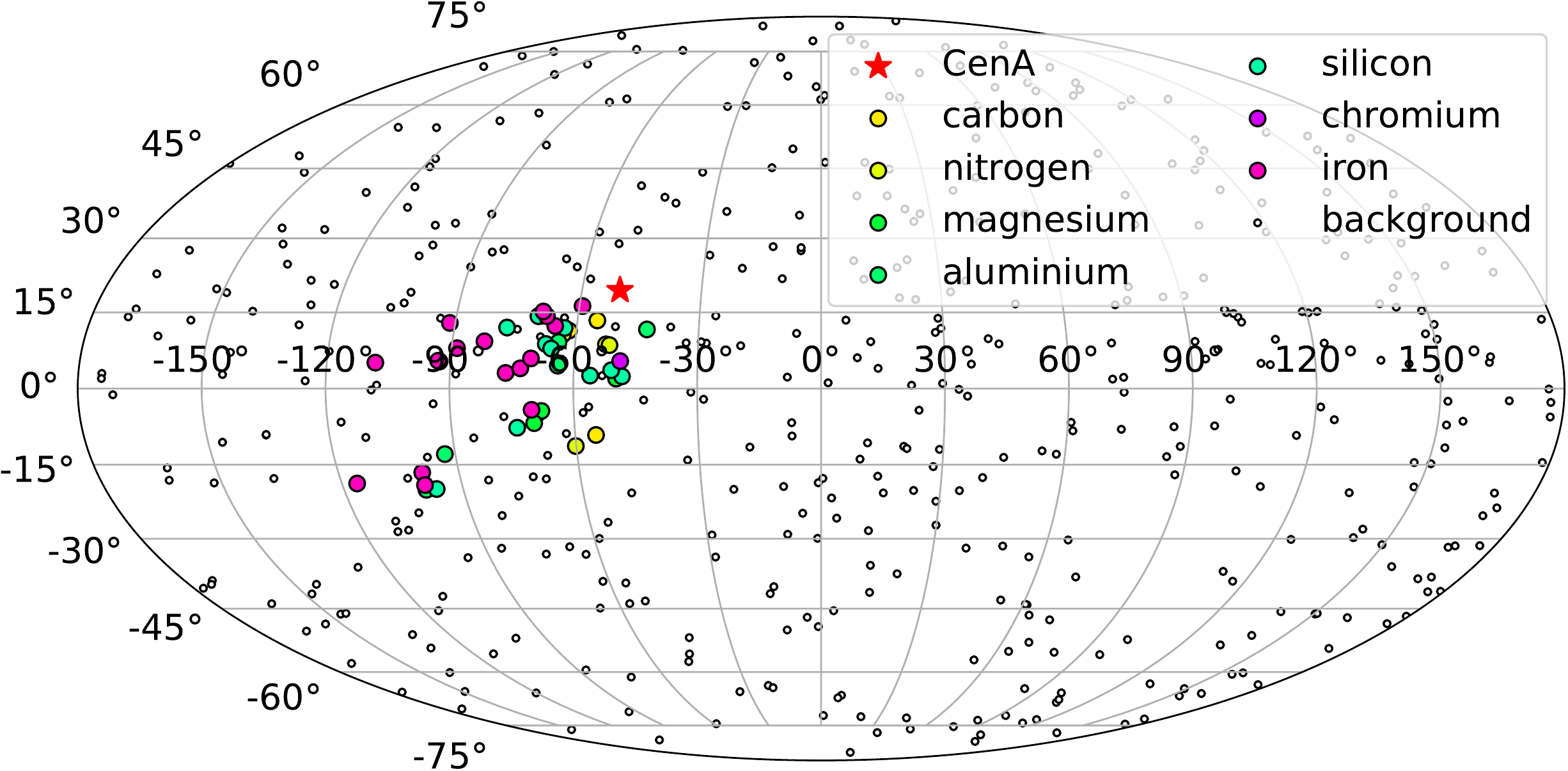}\\
	\figh{.20}{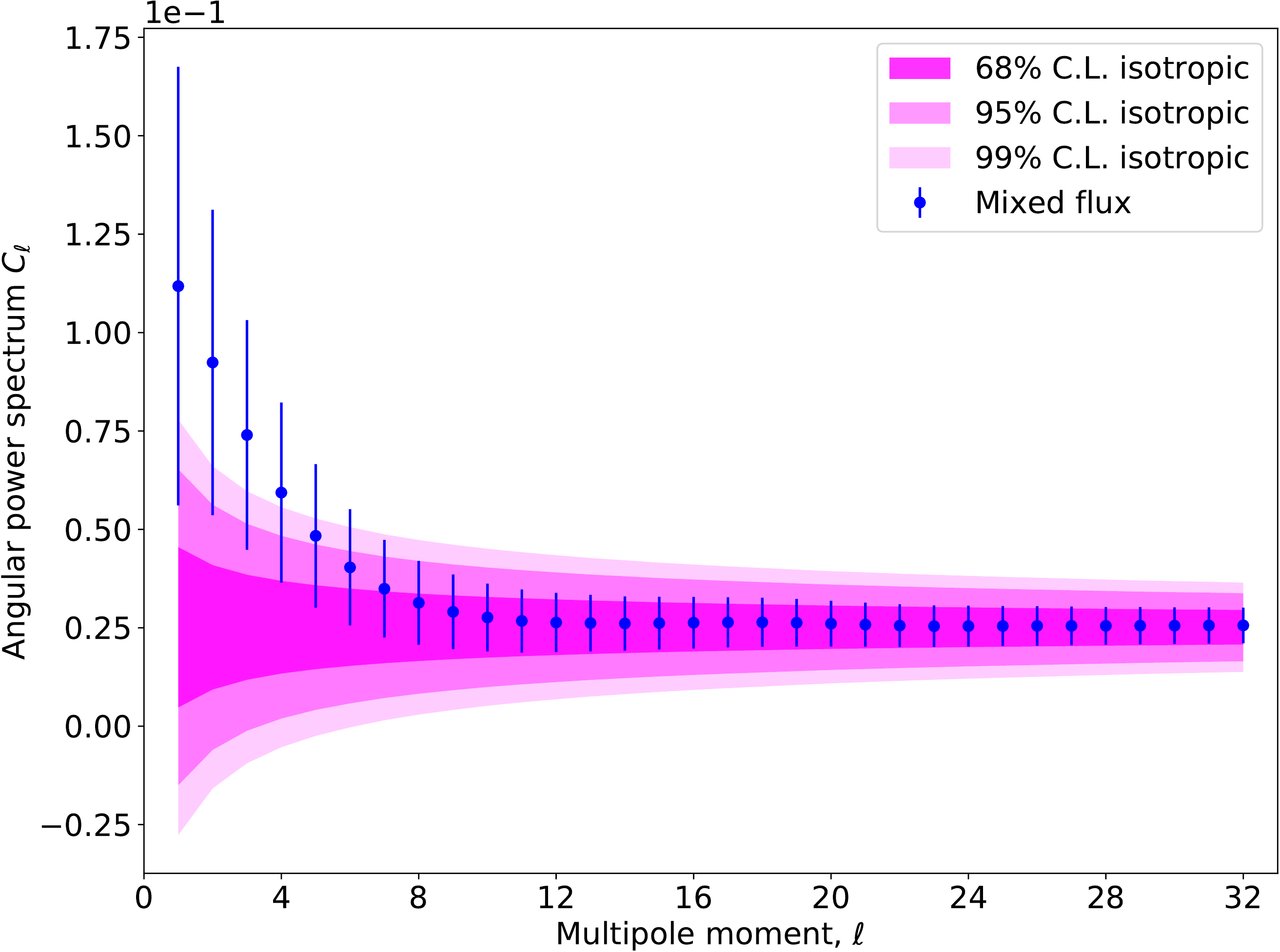}
	\figh{.194}{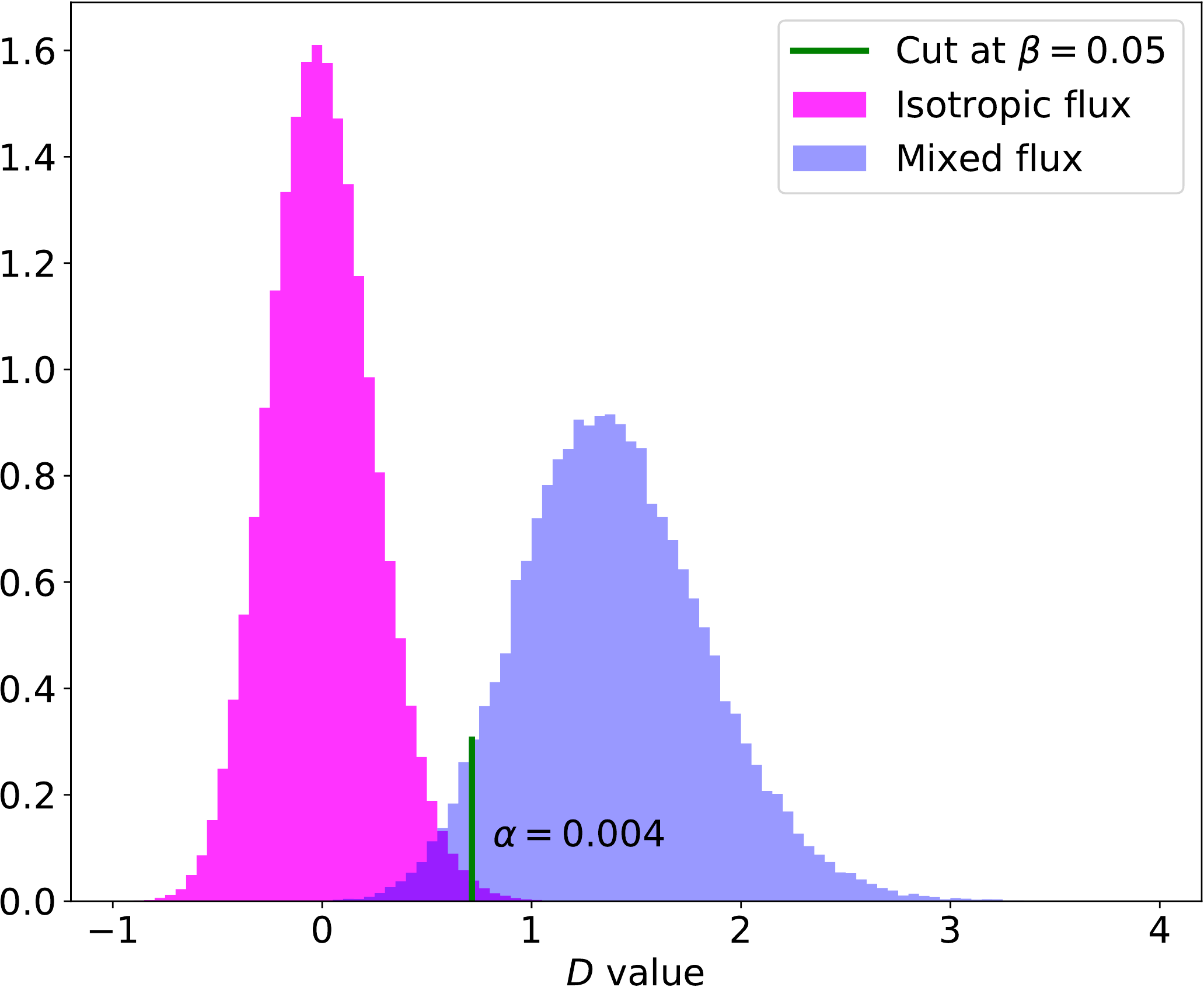}
	\caption{The case of $\Neecr=500$ and 9\% events coming from Cen~A.
	See the caption of Fig.~\ref{fig:NGC253} for other details.}
	\label{fig:CenA}
\end{figure}

\begin{figure}[!ht]
	\centering
	\figh{.21}{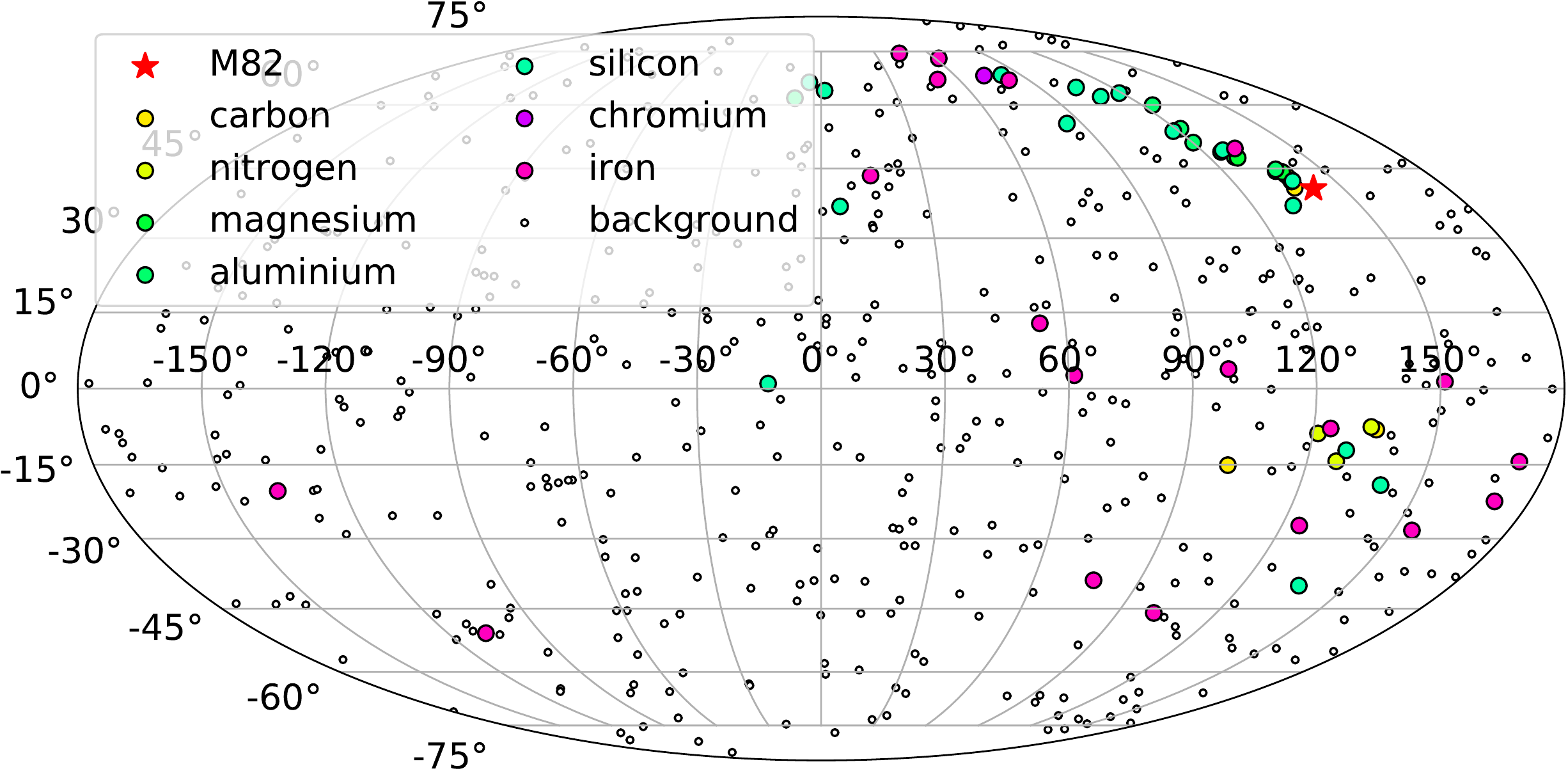}\\
	\figh{.20}{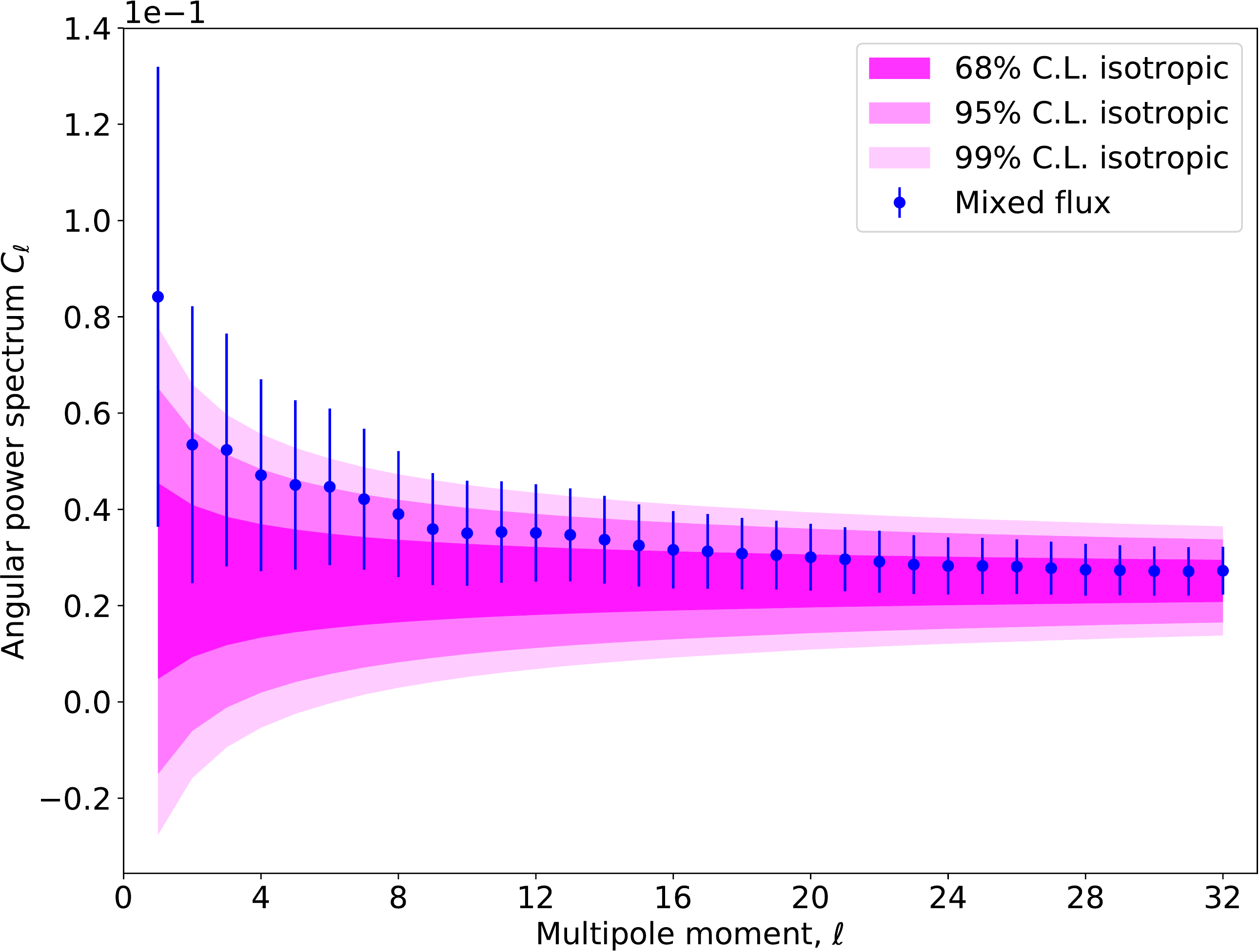}
	\figh{.194}{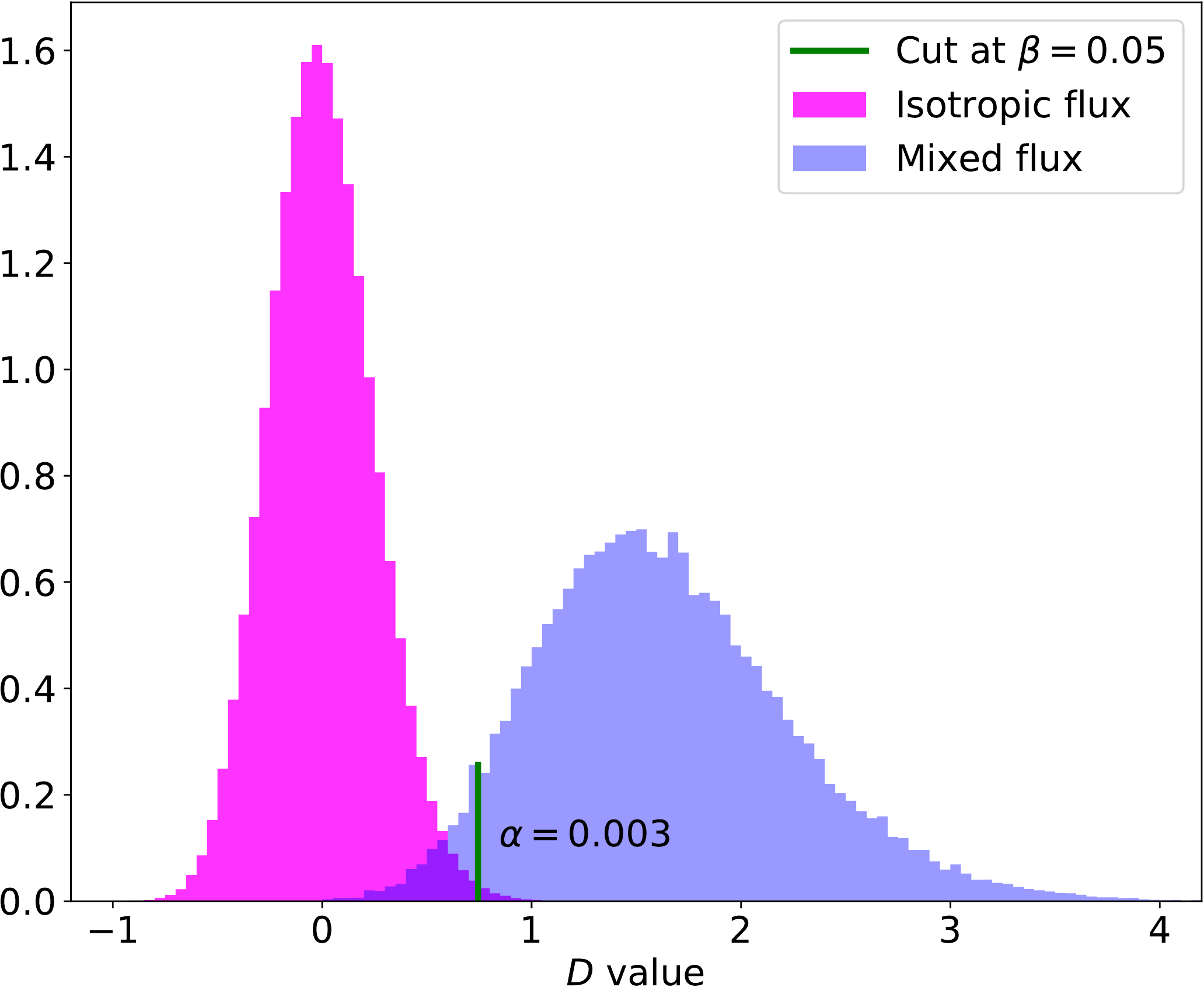}
	\caption{The case of $\Neecr=500$ and 11\% events coming from M82.
	See the caption of Fig.~\ref{fig:NGC253} for other details.}
	\label{fig:M82}
\end{figure}

\begin{figure}[!ht]
	\centering
	\figh{.21}{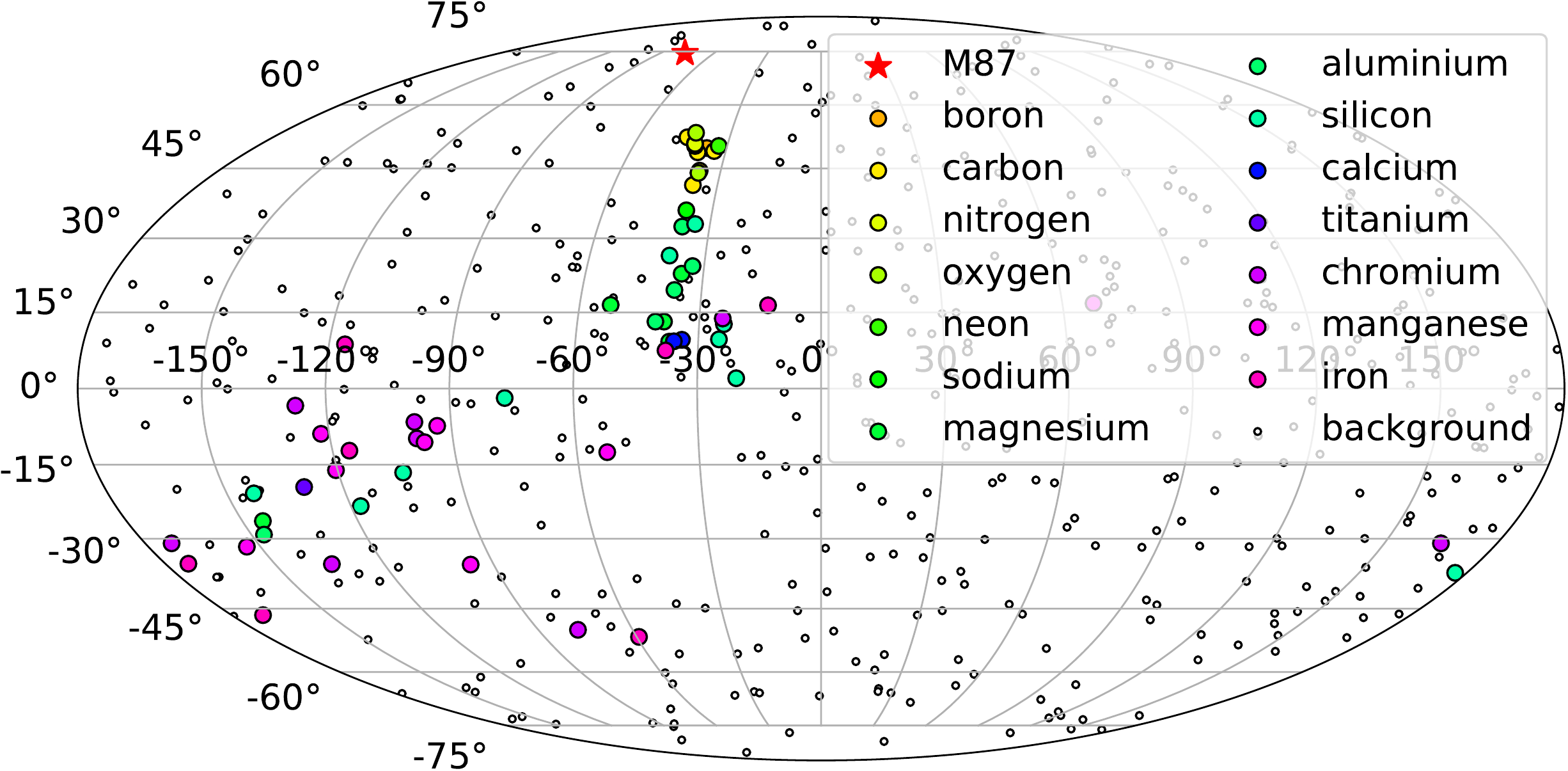}\\
	\figh{.20}{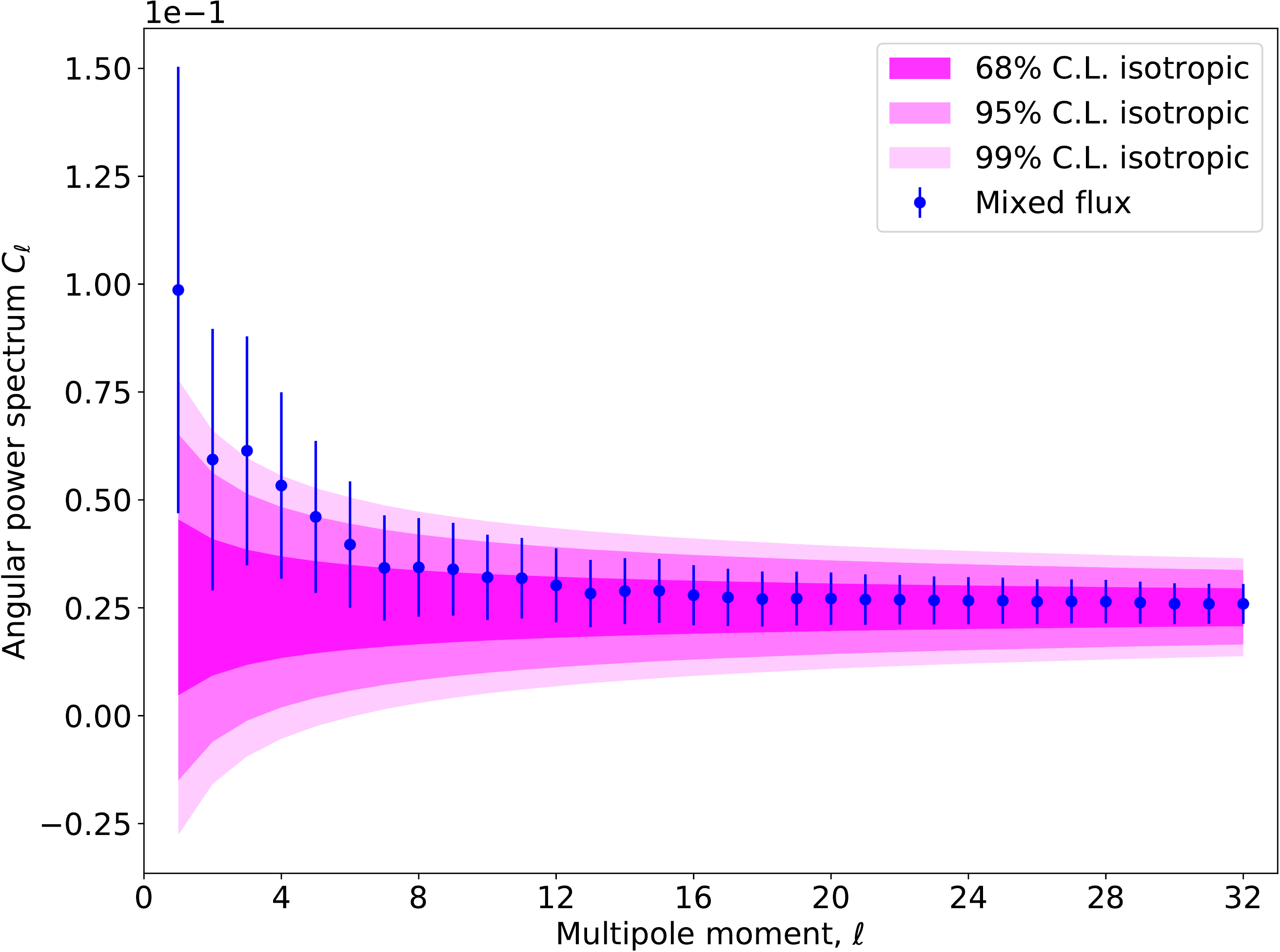}
	\figh{.194}{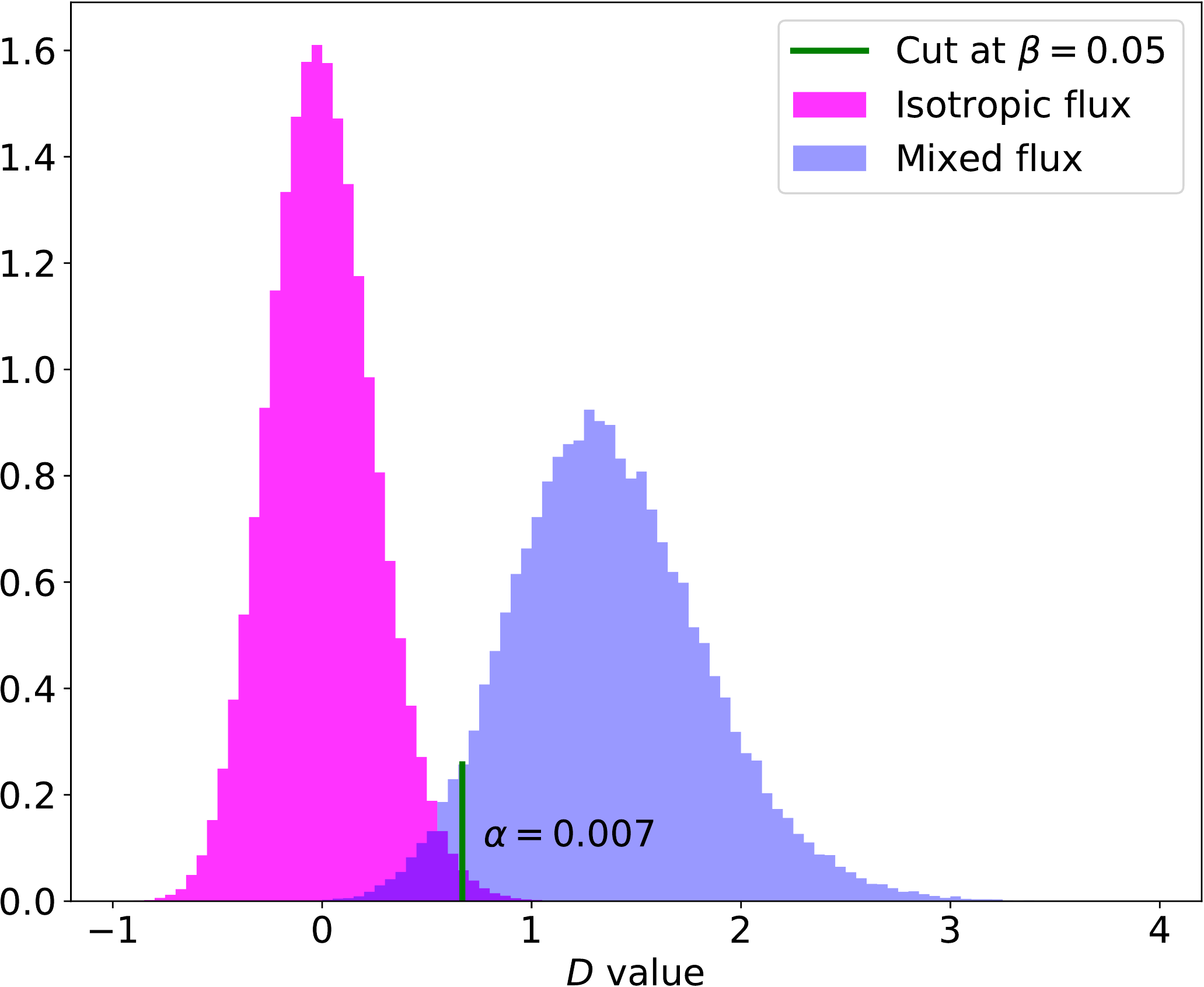}
	\caption{The case of $\Neecr=500$ and 12\% events coming from M87.
	See the caption of Fig.~\ref{fig:NGC253} for other details.}
	\label{fig:M87}
\end{figure}

\begin{figure}[!ht]
	\centering
	\figh{.21}{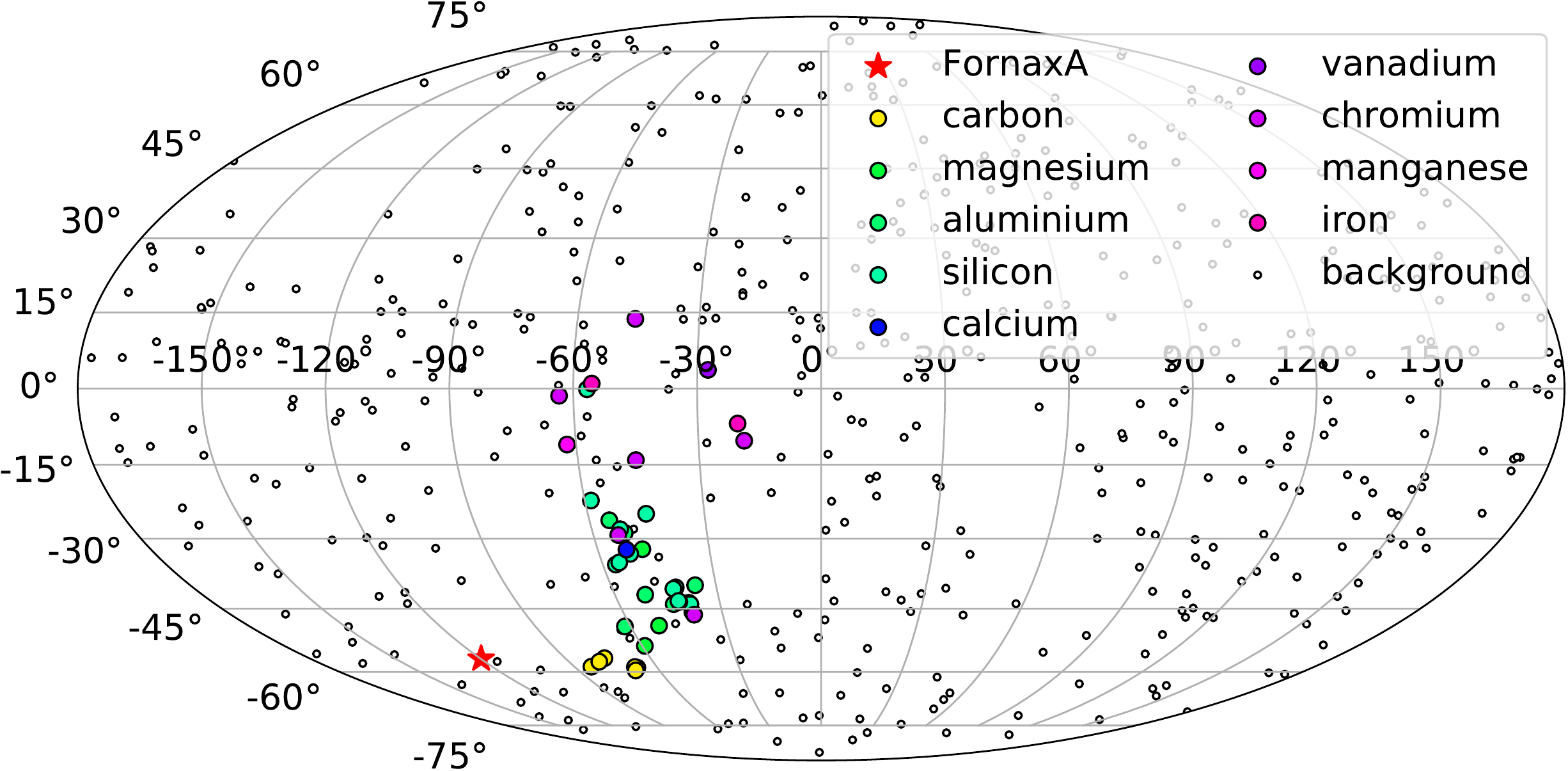}\\
	\figh{.20}{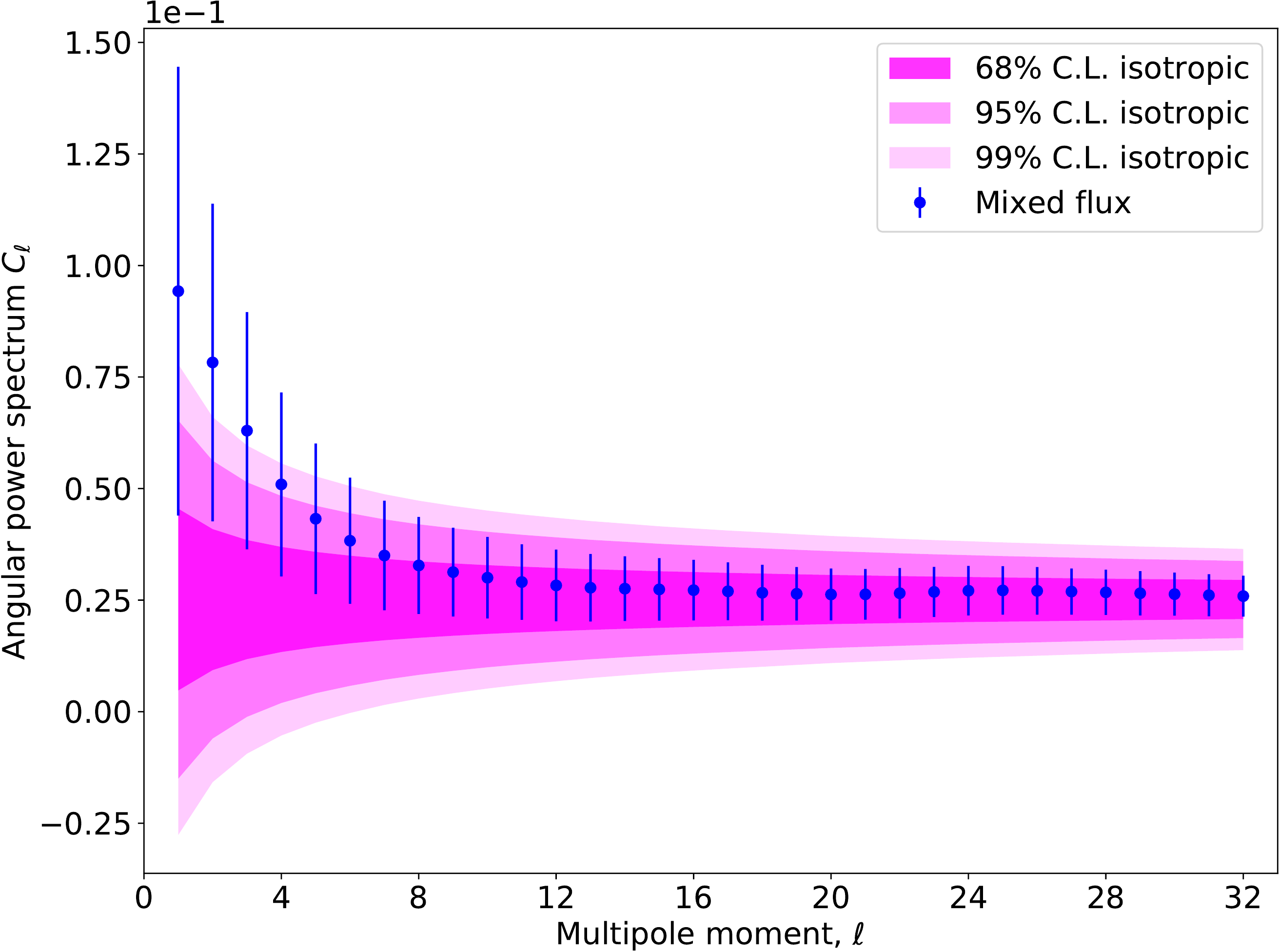}
	\figh{.194}{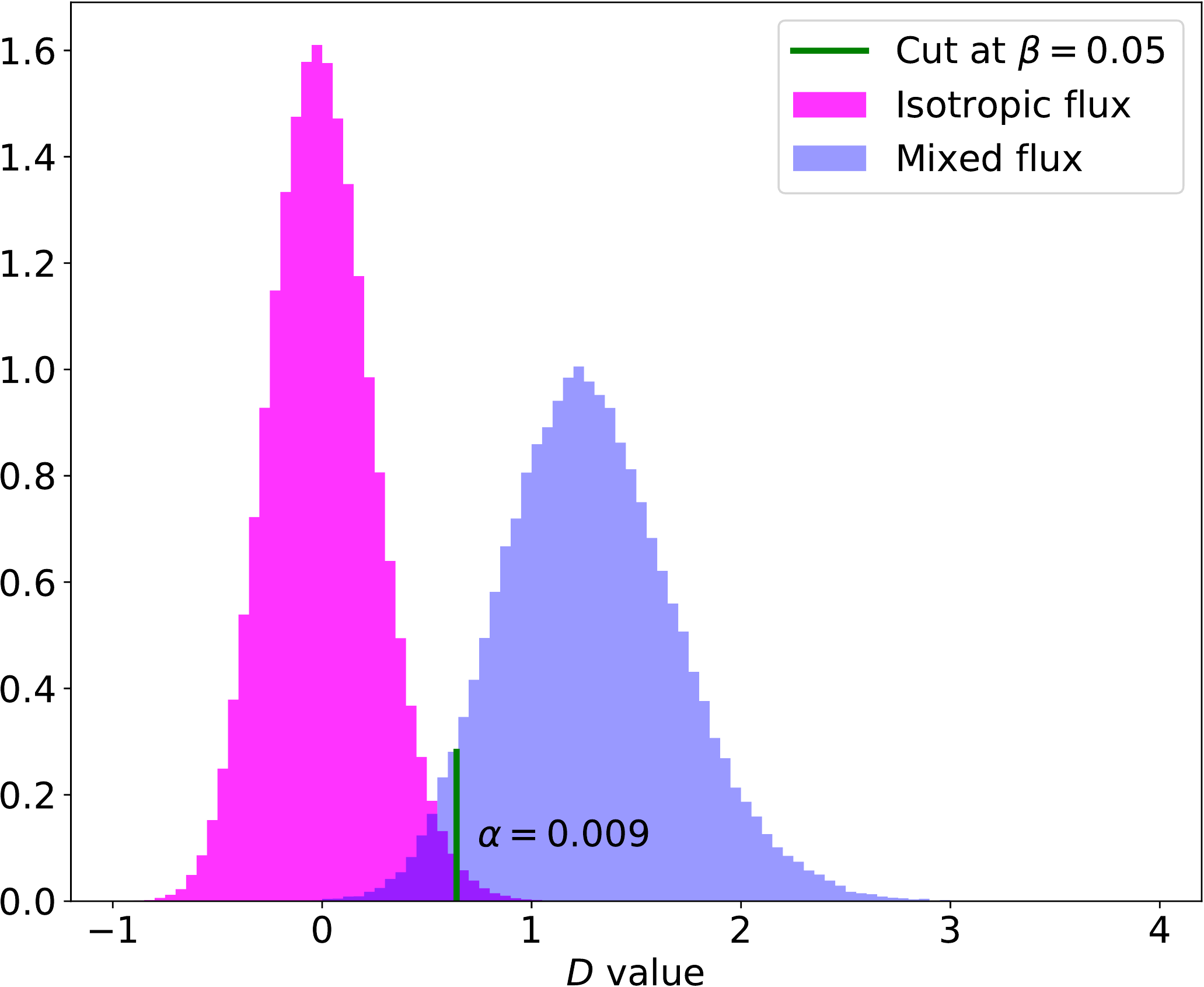}
	\caption{The case of $\Neecr=500$ and 8\% events coming from
	Fornax~A.
	See the caption of Fig.~\ref{fig:NGC253} for other details.}
	\label{fig:FornaxA}
\end{figure}

All results presented in Tables~\ref{table:percent}
and~\ref{table:alpha} were obtained for 500,000 isotropic samples
and at least 10,000 mixed samples for each~\Neecr.\footnote{%
    We tried up to $10^5$ mixed samples but simulations revealed
    that final results weakly depend on the size of samples beyond
    $10^4$.}
Results for $\Neecr=500$ are illustrated in
Figures~\ref{fig:NGC253}--\ref{fig:FornaxA} for each of the sources.
It is interesting to mention that a fraction of~\Diso\
greater than the median value of~\Dmix\ is $\lesssim10^{-5}$ in all
cases shown. 
Thus, the isotropy hypothesis will be rejected with a high
confidence level for a typical sample.

An important point to mention is the value of~\lmax\ used for
calculating the estimator~$D$ defined in Eq.~(\ref{eq:D}).
It became clear from simulations and can
be seen in Figures~\ref{fig:NGC253}--\ref{fig:FornaxA} that the
coefficients~\Cl\ calculated for mixed samples quickly converge to
isotropic values, so that it usually makes little sense to take into
account a contribution from multipoles with $\ell>16$.  All results in
terms of~$D$ presented here were obtained with one and the same
$\lmax=16$ for the sake of uniformity even it is not necessarily
optimal, see below.

Let us briefly comment on the numbers obtained for each of the sources
and presented in Tables~\ref{table:percent} and~\ref{table:alpha}.

The first three sources (NGC~253, Cen~A and
M82) are the closest among them. They are located at distances
$d\approx3.5$~Mpc~\cite{vanvelzen} at
different positions in the celestial sphere. Arrival directions of
nuclei coming from them demonstrate strikingly different patterns, see
the top panels in Figs.~\ref{fig:NGC253}--\ref{fig:M82}.
The behaviour of~\Cl\ is specific for each of the sources (see the
left bottom panels), and the more fuzzy is the pattern of arrival
directions of from-source UHECRs, the greater is their fraction in the
whole sample necessary to distinguish a large-scale anisotropy.
The percentage of from-source events required to reject the null
hypothesis with $\alpha\lesssim0.01$
varies in the range from~7\% to~11\% for $\Neecr=500$, 
and it grows with decreasing \Neecr\ approximately
inverse-proportionally to $\sqrt{\Neecr}$.
The latter is true for all the other sources.\footnote{We thank the
anonymous referee at JCAP for pointing this out.}

Next, M87 (Virgo~A) is an active galactic nucleus located in the
Northern hemisphere at $d\approx18.5$~Mpc from the Milky
Way~\cite{vanvelzen}.  It provides a specific pattern of
arrival directions of from-source UHECRs, distinct from the other
sources considered here, see Fig.~\ref{fig:M87}.
The pattern is comparatively fuzzy, resulting in a slightly higher
fraction of from-source events necessary to find an anisotropy.

Finally, the radio-loud Fornax~A (NGC~1316) galaxy is the most distant
source among those considered here, with $d\approx20$~Mpc.
A pattern of arrival direction of UHECRs coming from it is comparatively
compact resulting in mere~8\% of from-source events
among $\Neecr=500$ necessary to provide $\alpha=0.009$, see
Fig.~\ref{fig:FornaxA}.

Values of~$\alpha$ are sensitive to the fraction of
from-source events in samples with $\Neecr\gtrsim300$, so that
increasing their fraction by one percent can reduce~$\alpha$ by almost
an order of magnitude. This is especially pronounced for $\Neecr=500$.
For example, increasing the percentage of events coming from NGC~253 up
to 8\% for $\Neecr=500$ cuts down~$\alpha$ from $0.010$ down to
$\approx5\times10^{-4}$.

As was mentioned above, all presented results were obtained for $\lmax=16$
for the sake of uniformity of the analysis.  This does not mean this is
an optimal value for detecting a deviation from isotropic expectations
in each particular case since the behaviour of the~\Cl\ coefficients in
the angular power spectrum is specific for every source.  The
sensitivity of the estimator~$D$ in Eq.~(\ref{eq:D}) to deviations from
isotropy depends on the behaviour of~\Cl\ in each case.  For example,
the first kind error drops down to $\sim10^{-4}$ if one takes $\lmax=6$
for Cen~A with $\Neecr=300$.  On the other hand, it is advised to
increase~$\lmax$ slightly for M82 in order to obtain smaller~$\alpha$
due to a much slower decrease of the~\Cl\ coefficients for this AGN,
compare the left bottom panels in Figs.~\ref{fig:CenA}
and~\ref{fig:M82}.

\begin{figure}[!ht]
	\centering
	\figh{.32}{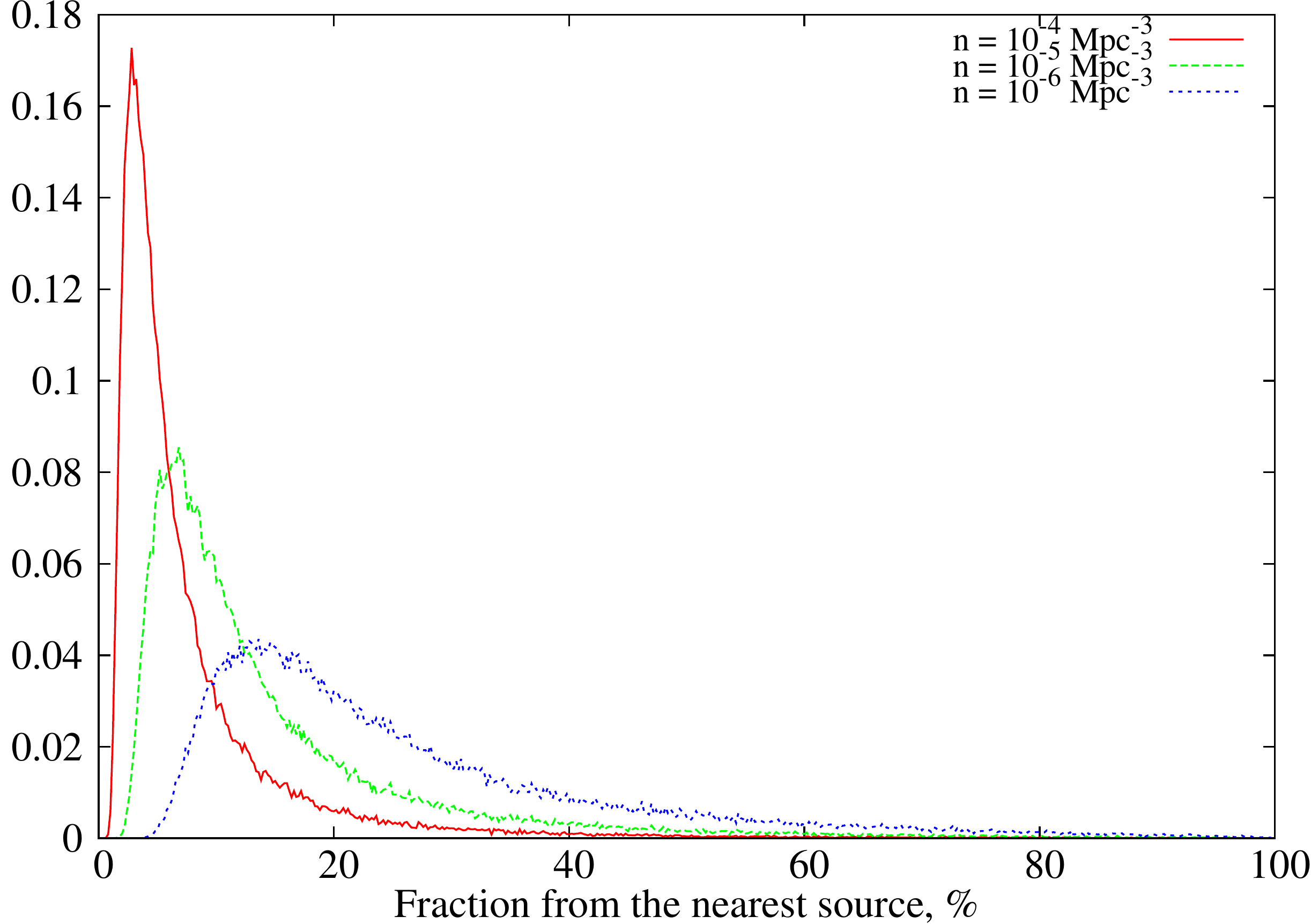}

	\caption{Distribution of contributions of the closest source
	in the total UHECR flux for different values of the source density
	$n=10^{-4},\, 10^{-5},\,10^{-6}$~Mpc$^{-3}$.}

	\label{fig:fraction_distro}
\end{figure}

It can be seen from Table~\ref{table:percent} that the threshold
value for the contribution from a single source that can be still detected
with a high confidence level is
around 10--15\% for $\Neecr\gtrsim300$, and it does not considerably depend
either on the source position on the celestial sphere or on the
distance to it. Thus, this value can be straightforwardly compared with
theoretical expectations. We used the simplest model of identical
sources uniformly distributed with a number density~$n$.  Given that the
characteristic path length~$L_c$ at the relevant energies is around
100~Mpc (see Fig.~\ref{fig:spectrum}), we have performed our simulations
in a $V_{\mathrm{box}}=(600~\mathrm{Mpc})^3$ box centered at the observer
position. Less than 5\% of the total flux comes from outside this  box, so we
have ignored that part. The total number of simulated sources was equal to
$N_{\mathrm{src}}=nV_{\mathrm{box}}$. The contribution from an
individual source located at a distance~$d$ was calculated as
\[
	\Phi=\exp(-d/L_c)/d^2.
\]
Finally, all contributions were summed up, and fractions of CRs from the
brightest (the closest in this set up) and the second-brightest sources
were calculated.  This procedure was repeated 100,000 times and that
gave a fairly good sampling of the distributions (see
Fig.~\ref{fig:fraction_distro}).  The median values of these distributions
for different values of density~$n$ are presented in
Table~\ref{table:fraction_distro}. It can be seen that the analysis of a
large-scale anisotropy observed by \keuso{} will have a potential to
constrain the density of identically distributed sources at the level
$n>(1-2)\times10^{-5}~\mathrm{Mpc^{-3}}$. This value should be compared
with the current limits coming from the non-observation of significant
clustering at intermediate angular scales in the same energy range by
the Pierre Auger Observatory: $n>6\times10^{-6}~\mathrm{Mpc^{-3}}$
\cite{PAO_density}.

\begin{table}[!ht]
	\caption{Percentage of UHECRs with $E>57$~EeV arriving from the
	closest and the second closest source in the setup of identical
	uniformly distributed sources with density~$n$.}
	\label{table:fraction_distro}
	\medskip
	\centering
	\begin{tabular}{|l|c|c|}
		\hline
		$n, \mathrm{Mpc}^{-3}$    &   Closest  &   Second closest\\
		\hline
		$10^{-4}$  &    5.2 &    1.8\\
		$3\times10^{-5}$  &    7.5 &    2.7\\
		$10^{-5}$  &    10.6 &    3.9\\
		$3\times10^{-6}$  &    15.0 &    5.0\\
		$10^{-6}$  &    20.9 &    6.3\\
		\hline
	\end{tabular}
\end{table}

\section{Anisotropy at lower energies}

It was suggested by Lemoine and Waxman~\cite{2009JCAP...11..009L} that
if anisotropy is found above some energy~$E$ and the composition
is assumed to be heavy at that energy (with nuclei of charge~$Z$), one
should also observe an even stronger anisotropy at energies above
rigidity $E/Z$ due
to the proton component of the flux emitted by the source that is
responsible for the observed anisotropy.
The statement was based on an assumption that the cosmic ray injection
spectrum at the source depends on rigidity only.

The idea was further developed in~\cite{2013ApJ...776...88L}.
It was argued in particular that if anisotropy is detected at
energies above $\sim60$~EeV and is caused by heavy nuclei, then an even
more significant anisotropy signal should be present at energies
close to the ankle due to the proton component.
The authors have also considered a case when protons are not injected
by the source. It was argued
that an anisotropy pattern may then occur at lower energies
due to the secondary proton signal produced by photodisintegration
of heavy nuclei.
However, the significance of an anisotropy predicted in this case at lower
energies is weaker than at higher ones unless the nearest source
is distant enough (typically beyond $30$~Mpc) and the maximal cosmic ray
injection energy $E_{\rm max} \gtrsim Z/26\times 10^3$~EeV
(see Fig.~2 in~\cite{2013ApJ...776...88L}).

The composition of UHECRs with energies above 57~EeV arriving
at Earth in the KKOS model is comparatively heavy, with $Z\ge4$.
It can be seen in the left panel of Fig.~\ref{fig:spectrum} that
the primary proton flux is strongly suppressed above $\sim3$~EeV in the
injection spectrum of the model.
It might be interesting to estimate if any large-scale anisotropy can be
observed at around the ankle of the CR energy spectrum
due to the secondary proton component present there,
even though this is below the energy threshold of the \keuso\
experiment, and an analysis of the original idea by Lemoine and Waxman
performed by Pierre Auger Collaboration did not reveal any
overdensities at lower energies in the regions where anisotropies
were found for energies above 55~EeV~\cite{2011JCAP...06..022P}.

\begin{figure}[!ht]
	\centering
	\figh{.34}{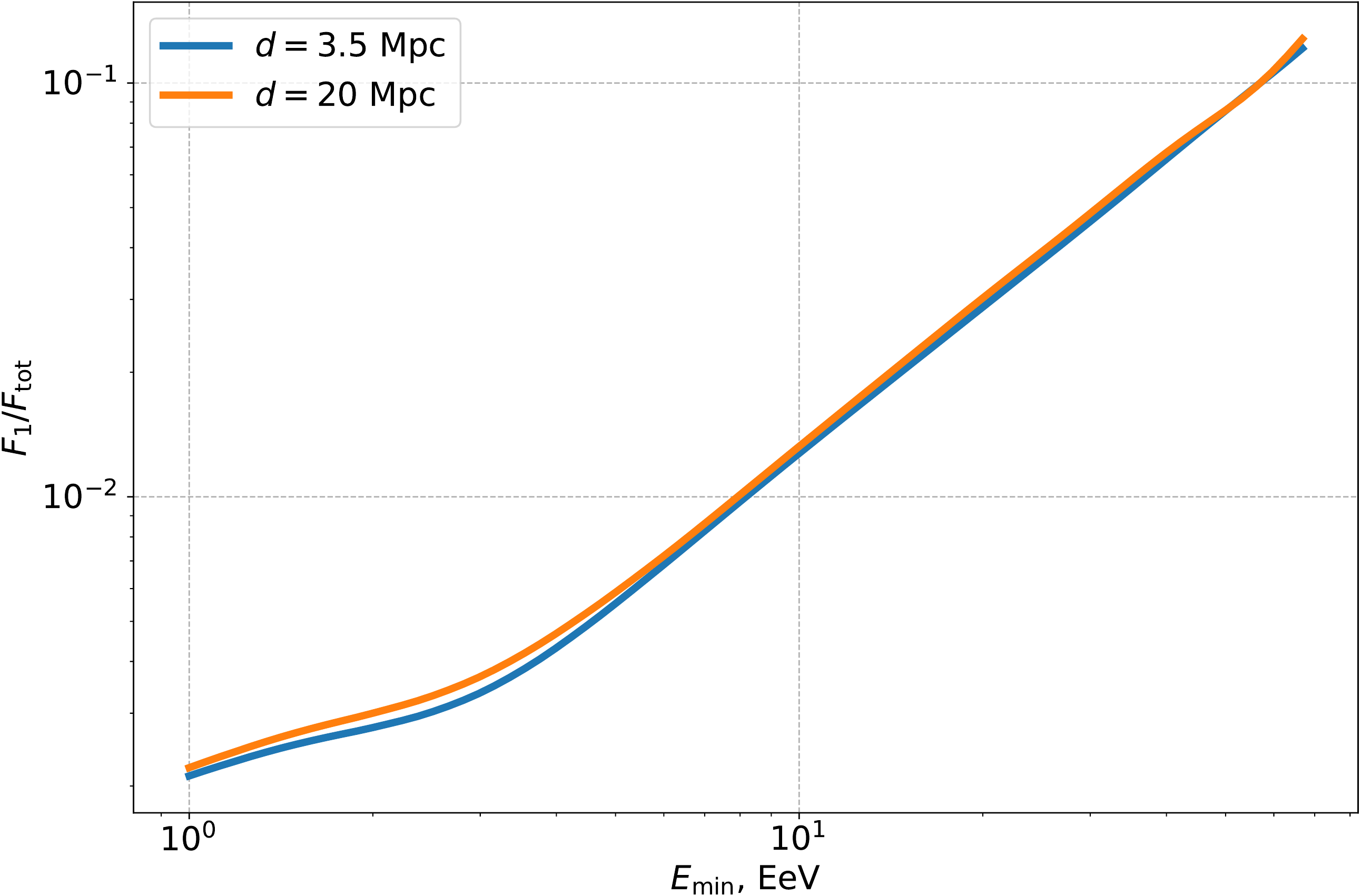}
	\caption{Dependence of the fraction of the
	flux~$F_1$ coming from a single source located at the given distance
	from the observer (3.5~Mpc and 20~Mpc) in
	the total incoming flux $F_\text{tot}$ above some energy
	threshold~$E_\mathrm{min}$. It is assumed that
	$F_1/F_\text{tot}=0.1$ for cosmic rays above $E_\mathrm{min}=57$~EeV.
	}
	\label{fig:frac}
\end{figure}

To address the question, we first evaluated how the fraction of
from-source events in the total flux scales with energy.  The dependence
of the fraction $F_1/F_\mathrm{tot}$ of cosmic rays arriving from a
single source in the total incoming flux on the minimal energy
$E_\mathrm{min}$ for distances $d=3.5$~Mpc and 20~Mpc is shown in
Fig.~\ref{fig:frac}.  For definiteness, the flux of UHECRs arriving from
a source is normalized in the figure so that 
$F_1/F_\mathrm{tot}=10\%$ for $E_\mathrm{min}=57$~EeV. The fraction grows
with energy since cosmic ray attenuation length drops with
increasing~$E$. 

It is clear from Fig.~\ref{fig:frac} that the fraction of
from-source events scales nearly linearly for $E_\mathrm{min}\gtrsim3$~EeV,
and there is very little difference between sources located at
distances 3.5~Mpc and 20~Mpc.
For both distances, the fraction equals approximately~1\% if one
considers energies above 8~EeV, which was the threshold value beyond
which a dipole anisotropy was detected by the Pierre Auger Collaboration
at more than a $5.2\sigma$ level of
significance~\cite{Auger-dipole-2017}.
(The amplitude of the dipole was found to be of $6.5^{+1.3}_{-0.9}$
percent, basing on the analysis of 32,187 events.)

\begin{figure}[!ht]
	\centering
	\figw{.48}{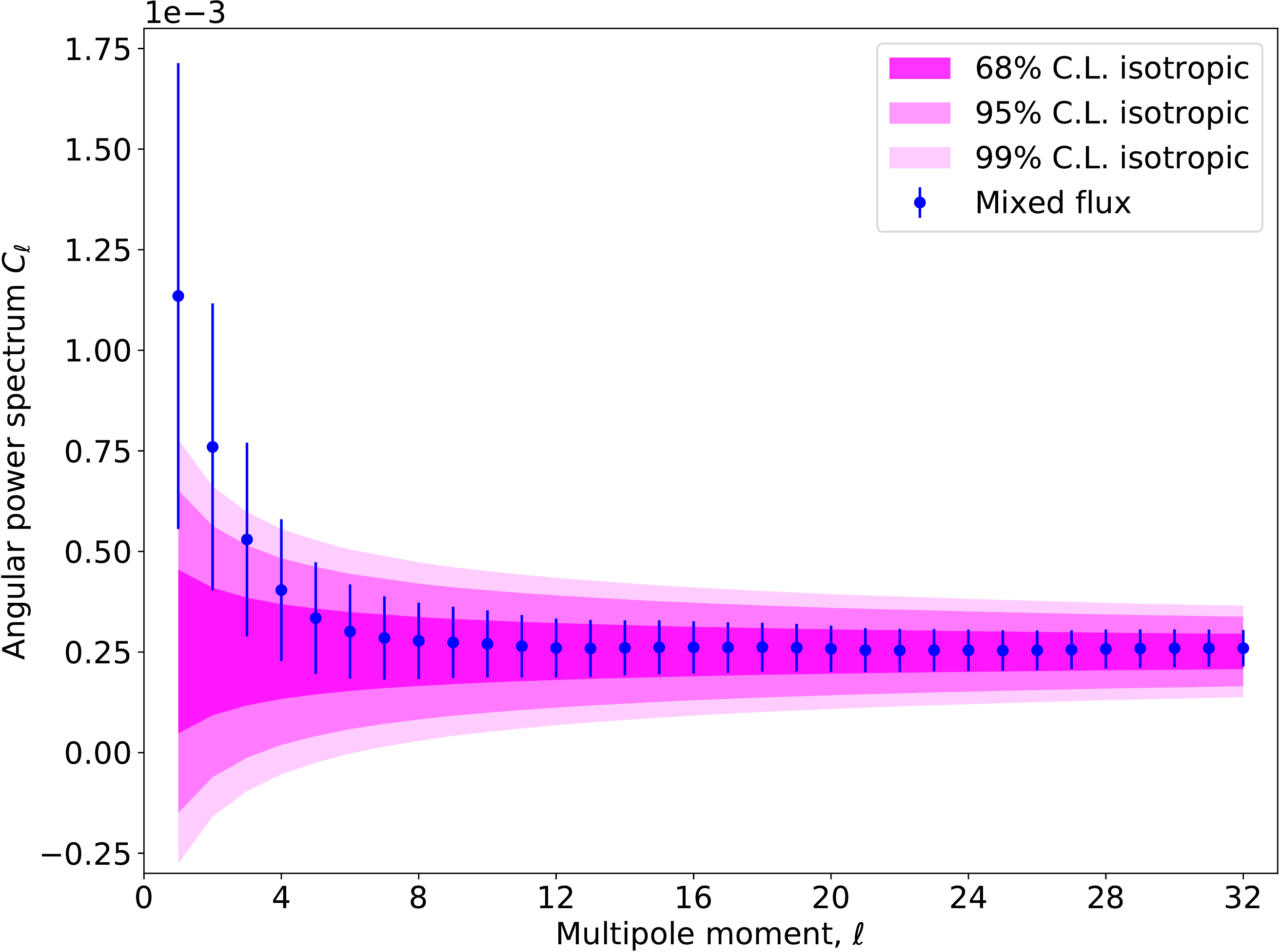}
	\figw{.48}{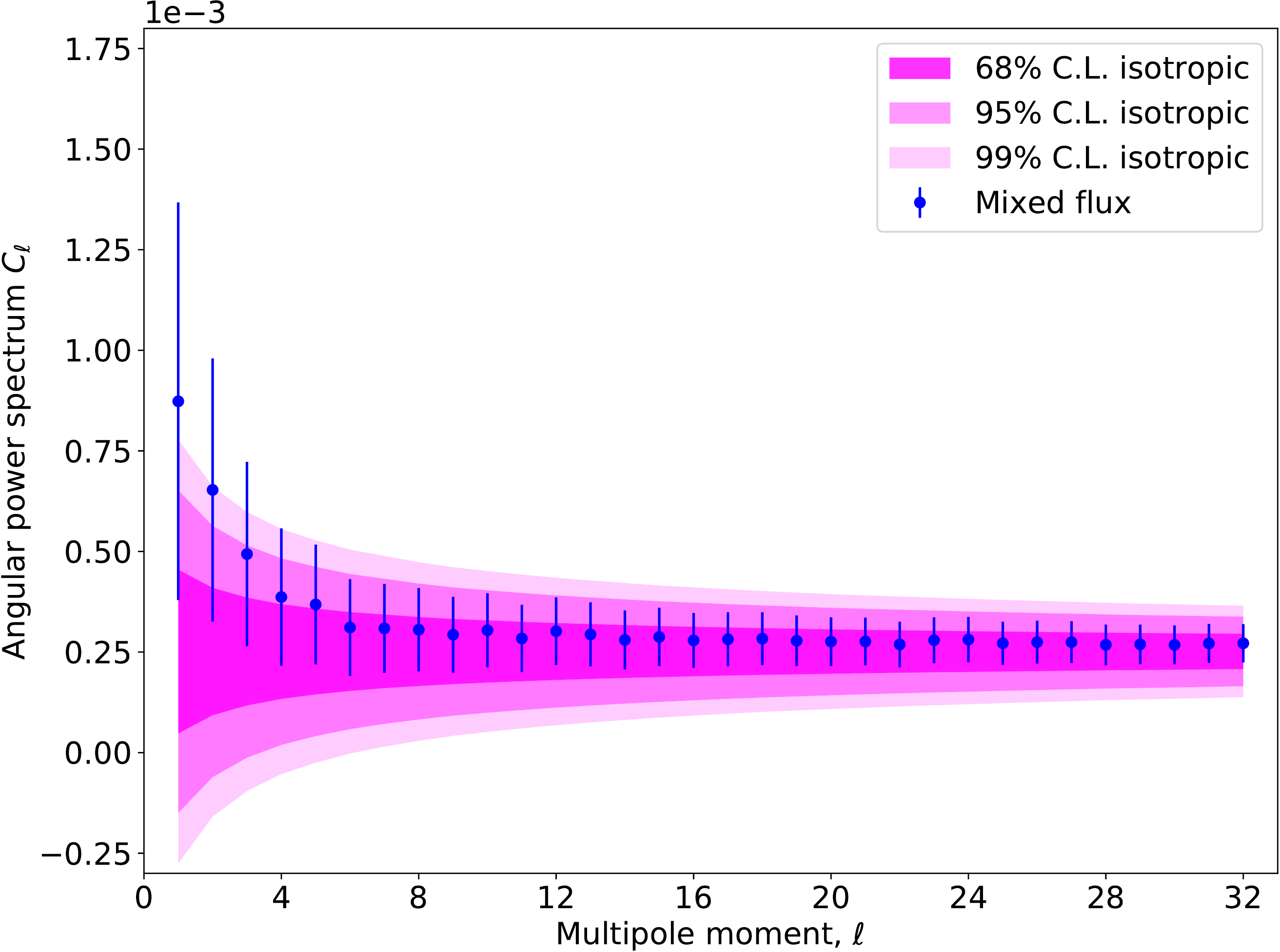}
	\caption{Angular power spectrum~\Cl\ for isotropic and mixed samples
	arriving from Cen~A (left) and Fornax~A (right) for cosmic rays 
	above 8~EeV and $F_1/F_\text{tot}=0.01$.
	Notation is explained in the caption of Fig.~\ref{fig:NGC253}.
	}
	\label{fig:8EeV}
\end{figure}

We performed simulations similar to those described above
for energies $\ge8$~EeV to check if a large-scale anisotropy
can be found in this
energy range assuming it is observed beyond 57~EeV.
We considered Cen~A and Fornax~A as candidates sources because
(i)~they are located at the opposite ends of the interval of distances,
(ii)~they demonstrate clearly different patterns of arrival
directions of UHECRs due to deflections in the Galactic magnetic field,
and (iii)~their positions are within the field of view of the Pierre Auger
Observatory.
The angular power spectra obtained for the two cases are shown in
Fig.~\ref{fig:8EeV}.
To make the plots, we simulated 500,000 isotropic samples and
50,000 mixed samples with from-source events comprising~1\%,
as suggested by Fig.~\ref{fig:frac}.
Each sample consisted of 50,000 events, which roughly corresponds to the
number of events used with $E\ge8$~EeV used in the analysis by
Auger~\cite{Auger-dipole-2017}.

It can be seen from Fig.~\ref{fig:8EeV} that a deviation from the
isotropic distribution is less pronounced in this case compared to
energies above 57~EeV, cf.\ Figs.~\ref{fig:CenA} and~\ref{fig:FornaxA}.
The first harmonic $C_1=(1.1\pm0.6)\times10^{-3}$
for Cen~A, and $C_1=(8.7\pm4.9)\times10^{-4}$ in the case of Fornax~A.
Since the dipole amplitude can be estimated as $\sqrt{9C_1/4\pi}$ (see,
e.g., \cite{2012MNRAS.427.1994G}), this immediately gives dipoles of
2.9\% and 2.5\% respectively, approximately 2.2--2.6 times less than the
dipole found by Auger~\cite{Auger-dipole-2017}.
An estimation obtained the same way in~\cite{Auger-aniso-2017}
with approximately 20,000 events gave an
amplitude of the dipole equal to $(6.0\pm1.5)\%$, thus more than two
times greater than those of ours.

The result qualitatively agrees with one of the conclusions
of~\cite{2019JCAP...01..018D} that the dipole amplitude increases at
higher energies.
It also explains why a dipole anisotropy near the ankle in the
CR energy spectrum associated with a nearby source has not been detected
by the current experiments assuming such an anisotropy exists at
energies beyond 57~EeV.

\section{Conclusions}

We have studied if the future \keuso\ orbital detector
will be able to observe
possible signatures of a large-scale anisotropy of
ultra-high-energy cosmic rays above 57~EeV arising
in the minimal model for extragalactic CRs and
neutrinos~\cite{Kachelriess:2017tvs}, a self-consistent scenario
attributing the origin of UHECRs and high energy neutrinos
to (possibly a subclass of) AGN.
We considered five possible candidate sources often discussed in
literature and allowed by the model (Centaurus~A, M82, NGC~253, M87 and
Fornax~A), focusing on the case of a single source providing a deviation
from an otherwise isotropic distribution.
Using extensive simulations performed with the 
publicly available, open-source
TransportCR and \mbox{CRPropa}~3 packages, we explored how anisotropies
depend on the energy threshold, the number of registered UHECRs
and the fraction of from-source events in the whole sample.
We demonstrate that an observation of $\gtrsim300$ events
above 57~EeV will allow
detecting a large-scale anisotropy with a high confidence level
providing the fraction of from-source events is $\simeq$10--15\%,
depending on a particular source, with a smaller source contribution
for larger samples.
We also show that anisotropy signatures originating from the same
nearby sources are not expected to be
strong at the ankle region in the KKOS model, which assumes
a heavy UHECR composition at the highest energies.

The presented results are generic and can be applied to other
future experiments with a full-sky coverage and a uniform exposure,
including the planned POEMMA mission~\cite{POEMMA}.

\acknowledgments

We thank the anonymous referee for numerous insightful comments on the
earlier versions of the manuscript.
This research has made use of the NASA/IPAC Extragalactic Database
(NED), which is operated by the Jet Propulsion Laboratory, California
Institute of Technology, under contract with the National Aeronautics
and Space Administration, and of the SIMBAD database, operated at CDS,
Strasbourg, France~\cite{simbad}.
We have employed IPython~\cite{ipython} to perform calculations
and Matplotlib~\cite{matplotlib} to make figures.
Some of the results in this paper have been derived using the HEALPix
package~\cite{healpix}.
The work was done with partial financial support from the Russian
Foundation for Basic Research grant No.\ 16-29-13065.
MP acknowledges the support from the Program of development of
M.\,V.~Lomonosov Moscow State University (Leading Scientific School
``Physics of stars, relativistic objects and galaxies'').

\bibliographystyle{JHEP}
\bibliography{aniso}

\end{document}